\documentclass[12pt,preprint,floatfix]{aastex}

\shorttitle{Stellar Collisions and UCXBs}
\shortauthors{Lombardi et al.}
\tighten 

\newcommand\vp[1]{\vphantom{#1}}

\begin{document}
\title{Stellar Collisions and Ultracompact X-ray Binary Formation}
\author{J.~C.~Lombardi,~Jr.$^1$, Z.~F.~Proulx$^1$, K.~L.~Dooley$^1$,
E.~M.~Theriault$^1$, N.~Ivanova$^2$, and F.~A.~Rasio$^2$}
\altaffiltext{1}{Department of Physics and Astronomy, Vassar
College, 124 Raymond Avenue, Poughkeepsie, NY 12604-0745}
\altaffiltext{2}{Department of Physics and
Astronomy, Northwestern University, Evanston, IL 60208}

\begin{abstract}
We report the results of new smoothed particle hydrodynamics (SPH)
calculations of parabolic collisions between a subgiant or slightly
evolved red-giant star and a neutron star (NS). Such collisions are
likely to provide the dominant formation mechanism for ultracompact
X-ray binaries (UCXBs) observed today in old globular clusters.  In
particular, we compute collisions of a $1.4\,M_\odot$ NS with
realistically modelled parent stars of initial masses $0.8$ and
$0.9\,M_\odot$, each at three different evolutionary stages
(corresponding to three different core masses $m_c$ and radii $R$).
The distance of closest approach for the initial orbit varies from
$r_p = 0.04\,R$ (nearly head-on) to $r_p = 1.3\,R$ (grazing). These
collisions lead to the formation of a tight binary, composed of the NS
and the subgiant or red-giant core, embedded in an extremely diffuse
common envelope (CE) typically of mass $\sim 0.1  $ to $0.3\,M_\odot$.
Our calculations follow the binary for many hundreds of orbits,
ensuring that the orbital parameters we determine at the end of the
calculations are close to final. Some of the fluid initially in the
giant's envelope, from 0.003 to $0.023\,M_\odot$ in the cases we
considered, is left bound to the NS. The eccentricities of the
resulting binaries range from about 0.2 for our most grazing collision
to about 0.9 for the nearly head-on cases. In almost all the cases we
consider, gravitational radiation alone will cause sufficiently fast
orbital decay to form a UCXB within a Hubble time, and often on a much
shorter timescale.  Our hydrodynamics code implements the recent SPH
equations of motion derived with a variational approach by Springel \&
Hernquist and by Monaghan.  Numerical noise is reduced by enforcing an
analytic constraint equation that relates the smoothing lengths and
densities of SPH particles.  We present tests of these new methods to
help demonstrate their improved accuracy.
\end{abstract}


\section{Introduction and Motivation}

Ultracompact X-ray binaries (UCXBs) are bright X-ray sources, with
X-ray luminosities $L\gtrsim 10^{36}$~erg~s$^{-1}$, and very short
orbital periods $P \lesssim 1$~hr.
%
%
UCXBs
are generally believed to be powered by accretion from a white dwarf
(WD) donor onto a neutron star (NS), with the mass transfer driven
by gravitational radiation \citep{1987ApJ...322..842R}.
The abundance of bright X-ray binaries in
Galactic globular clusters exceeds that in the field by many orders
of magnitude, indicating that these binaries are formed through
dynamical processes \citep{Clark}. Indeed the stellar encounter rate
in clusters has been shown to correlate with the number of close
X-ray binaries \citep{2003ApJ...591L.131P}. In this paper, we use
SPH calculations to investigate how UCXBs can be formed through
direct physical collisions between a NS and a subgiant or small red
giant \citep{V87,dbh92,rs91,apjletter}. In
this scenario, the collision strips the subgiant or red giant,
leaving its core orbiting the NS in an extremely diffuse common
envelope (CE) of residual gas. Through the combined dissipative
effects of CE evolution and, more importantly, gravitational
radiation, the orbit decays until the core of the stripped giant
overflows its Roche lobe and mass transfer onto the NS begins.

Many recent studies of UCXBs have revealed their important role in a
number of different contexts. They may be dominant in
the bright end of the X-ray luminosity function in
elliptical galaxies \citep{BD2004}. UCXBs allow us to better
understand in general the stellar structure and evolution of
low-mass degenerate or quasi-degenerate objects \citep{DB2003,2005ApJ...624..934D}.
They are also connected in a fundamental way to NS recycling and
millisecond pulsar formation. Indeed, three out of five
accretion-powered millisecond X-ray pulsars known in our Galaxy are
UCXBs \citep{C2005}.
Finally, UCXBs may likely be the progenitors of the many eclipsing
binary radio pulsars with very low-mass companions observed in
globular clusters \citep{RPR2000,F2005}.

Several possible dynamical formation processes for UCXBs have been
discussed in the literature. Exchange interactions between NSs and
primordial binaries provide a natural way of forming possible
progenitors of UCXBs \citep{RPR2000}. This may well dominate the
formation rate when integrated over the entire dynamical history of
an old globular cluster. However, it is unlikely to be significant
for bright UCXBs observed today. This is because the progenitors
must be intermediate-mass binaries, with the NS companion massive
enough for the initial mass transfer to become dynamically unstable,
leading to CE evolution and significant orbital decay. Instead, all
MS stars remaining today in a globular cluster (with masses below
the MS turn-off mass $m_{\rm to}\simeq 0.8\,M_\odot$) have masses
low enough to lead to {\em stable\/} mass transfer (and the
formation of wider LMXBs with non-degenerate donors). Alternatively,
some binaries with stable mass transfer could evolve to ultra-short
periods through magnetic capture
\citep{PS88,PRP2002}. However, producing UCXBs through this type of
evolution requires extremely efficient magnetic braking
and a very careful tuning of initial orbital period and donor mass, and it
is therefore very unlikely to explain most sources
\citep{2005A&A...431..647V,Marcspreprint}.

\begin{figure}
\plotone{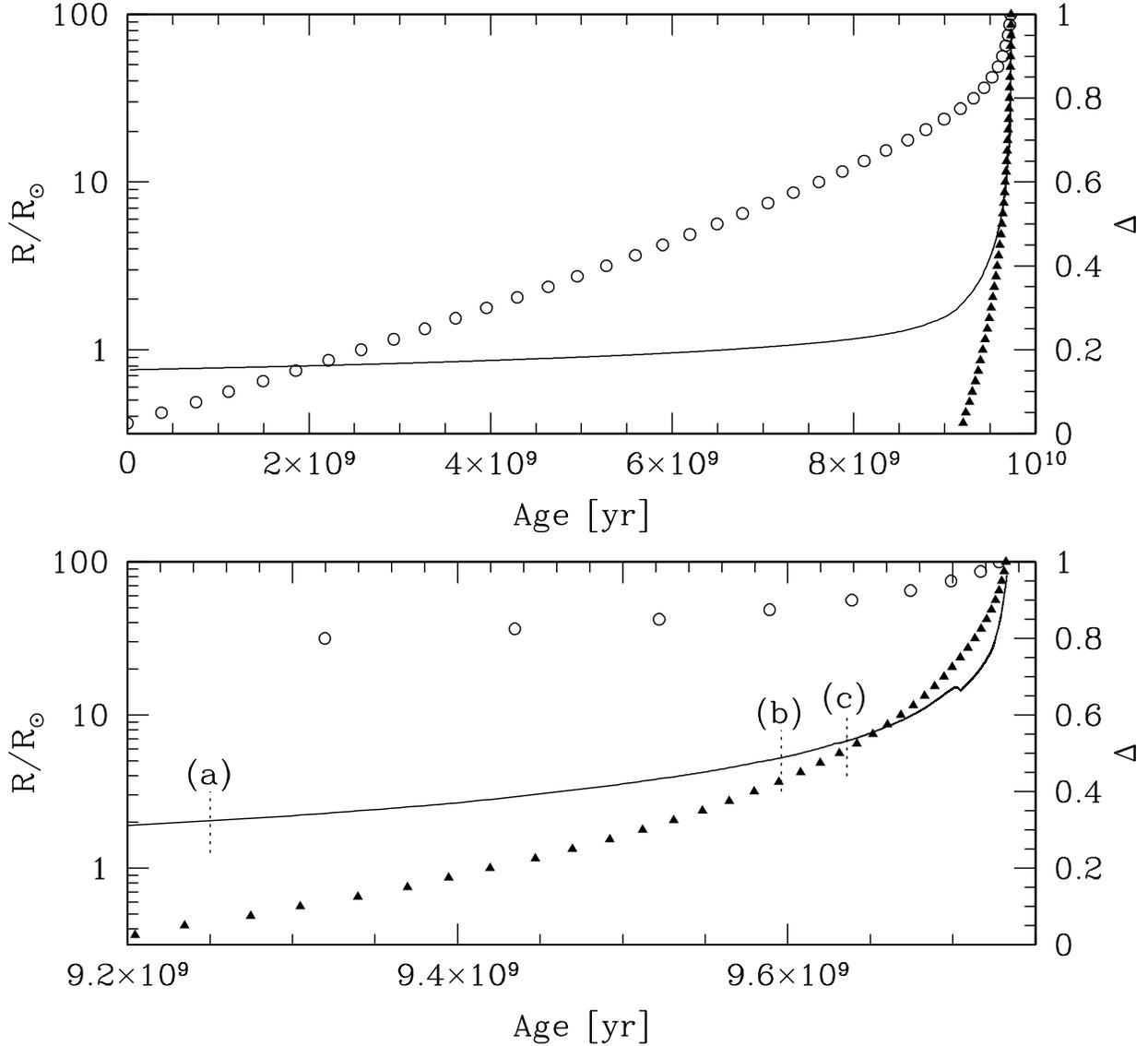}
\caption{Normalized number of physical collisions
$\Delta$ shown for a star of mass $0.9 M_\odot$ in a cluster with
$v_\infty=10$ km s$^{-1}$.  The upper panel shows the evolution from
the zero age main sequence to the top of the giant branch, and the
lower panel zooms in to show the subgiant and red giant stages. The
solid curve represents the stellar radius $R$ (left axis). Symbols show
$\Delta$ (right axis) at the current age: open circles are for a
$t_0$ corresponding to the zero age main sequence,
while solid triangles are for a $t_0$ corresponding to the terminal
age main sequence. The markers (a), (b), and (c) in the bottom panel
correspond to three of the stars modelled in this paper, namely
SG0.9a, RG0.9b, and RG0.9c respectively (see \S \ref{init_data}).
\label{coll}}
\end{figure}

Figure \ref{coll} helps provide the motivation for our consideration
of collisions with subgiants or small red giants by showing that a
significant fraction of collisions occur during these stages.  This
figure displays, for a star of mass $0.9 M_\odot$, the stellar radius
$R$ as a function of time as well as the normalized number of physical
collisions $\Delta (\tau) = \int_{t_0}^{\tau} r(t)dt/\int_{t_0}^{t_f}
r(t)dt$, where $t_0$ is the initial time of consideration, $t_f$
corresponds to the end of the red giant stage, $r(t)$  is the rate of
collisions, $r(t)\propto \sigma(t)\propto R^2 \left[1+2G(M+M_{\rm
NS})/(Rv_\infty^2)\right]$, $\sigma$ is the cross-section for the
physical collision with a NS, $M_{\rm NS}$ is the mass of the NS, and
$v_\infty$ is the relative velocity at infinity.  From the open
circles in this figure we see, for example, that nearly 80\% of the
stars with $M\approx 0.9 M_\odot$ that do collide will suffer their
collision before leaving the main sequence (at $t=9.2\times 10^9$
yr). This leaves more than 20\% of such stars to collide while a
subgiant or red giant.  Of these, more than 60\% collide while the
star still has a radius $R<10 R_\odot$ (see the triangular data point
at the time when the radius $R=10R_\odot$).  Therefore, due to their
being larger than main sequence stars and evolving less rapidly than
large red giants, subgiants and small red giants are therefore
significant participants in stellar collisions.

In a previous paper, we
showed that direct physical collisions between NSs and 
subgiants or small red giants
can provide a sufficient formation rate to explain the observed numbers
of UCXBs \citep{apjletter}.  In this paper, we present in detail
the hydrodynamics calculations of such collisions.
Our paper is organized as follows. In \S\ref{num_methods}, we
present the SPH method. In \S\ref{comparison}, we present results
from one- and three-dimensional test simulations that compare the
performance of the variational and classical SPH equations of motion.  In
\S\ref{init_data}, we present the SPH models of the parent stars we
consider, and in \S\ref{results} these models are used in collision
simulations.  Finally in \S\ref{future}, we summarize our results
and discuss future work.

\section{Numerical Methods\label{num_methods}}

SPH is the most widely used hydrodynamics scheme in the astrophysics
community. It is a Lagrangian particle method, meaning that the fluid is
represented by a finite number of fluid elements or ``particles.''
Associated with each particle $i$ are, for example, its position
${\bf r}_i$, velocity ${\bf v}_i$, and mass $m_i$.  Each particle
also carries a purely numerical smoothing length $h_i$ that
determines the local spatial resolution and is used in the
calculation of fluid properties such as acceleration and density.
See \citet{ras99} for a review of SPH, especially in the context of stellar collisions.

Not surprisingly, there are many variations on the SPH theme.  For
example, one can choose to integrate an entropy-like variable,
internal energy, or total energy.  There are also different ways to
estimate pressure gradient forces and a variety of popular choices
for the artificial viscosity. Tests and comparisons of many of these
schemes are presented in \citet{lsrs99}.

The classical formulations of SPH have proven more than adequate for
numerous applications, but because their underlying equations of
motion implicitly assume that $h_i$ is constant in time, they do not
simultaneously evolve energy and entropy strictly correctly when
adaptive smoothing lengths are used \citep{ras91,her93}.  In
cosmological SPH simulations, for example, the resulting error in
entropy evolution can significantly affect the final mass
distribution \citep{2003ApJ...597..878S}. Seminal papers addressing
the problem \citep{np93,np94,setal96} allowed for adaptive smoothing
lengths, but in a somewhat awkward manner that was not generally
adopted by the SPH community.

More recently, \citet{sh02} and \citet{mon02} have used a
variational approach to derive the SPH equations of motion in a way
that allows very naturally for variations in smoothing length.
As we show in \S \ref{comparison}, this formalism works
especially well when coupled with a new approach in which $h_i$ and
$\rho_i$ are solved for simultaneously \citep{mon02}.
The idea is that some
function of the particle density and smoothing length is kept constant
(see \S\ref{subsection:density}), thereby satisfying
the requirement of the variational derivation that the smoothing lengths
be a differentiable function of particle positions. The purpose
of this section is to present our implementation of these techniques.

\subsection{Density and Smoothing Length\label{subsection:density}}

An estimate of the fluid density at ${\bf r}_i$ is calculated from
the masses and positions of neighboring particles
as a local weighted average,
\begin{equation}
\rho_i=\sum_j m_j W_{ij}(h_i)\,,
\label{rho}
\end{equation}
where $W_{ij}(h_i)=W(r_{ij},h_i)$ is a smoothing (or interpolation)
kernel with $r_{ij}\equiv \left|{\bf r}_i-{\bf r}_j\right|$. We use
the second-order accurate kernel of \citet{Monaghan-Lattanzio1985},
\begin{equation}
W(r,h)={1\over\pi h^3}
\cases{
  \medskip\displaystyle
  1-{3\over2}\left({r\over h}\right)^2+
    {3\over4}\left({r\over h}\right)^3,
    &$\displaystyle0\le{r\over h}<1$\,;\cr\medskip
  \displaystyle{1\over4}\left(2-{r\over h}\right)^3,
    &$\displaystyle1\le{r\over h}<2$\,;\cr
  0\,,
    &$\displaystyle{r\over h}\ge2$\,.}
\end{equation}

Because the variational approach of deriving the SPH equations of motion
requires that the smoothing lengths be a function of the particle
positions, we choose an analytic function $H_i(\rho)$ and
solve $h_i=H_i(\rho_i)$ simultaneously with equation~(\ref{rho});
this idea is presented in \citet{mon02} where it is credited to
Bonet.  Note that this solution can be found one particle at a time,
as equation~(\ref{rho}) depends only on the smoothing length $h_i$
and not on any $h_j$ ($j\ne i$).
The analytic function we use is
\begin{equation}
H_i(\rho)=\left(h_{{\rm max},i}^{-1} + b_i \rho^{1/\nu}\right)^{-1}\,, \label{Hofrho}
\end{equation}
where $\nu$ is the number of dimensions, $h_{{\rm max},i}$ is the
maximum allowed smoothing length for particle $i$ (the value
approached as the density $\rho$ tends to zero), and $b_i$ is a free
parameter that can be adjusted according to the desired initial
number of neighbors.  For our one-dimensional test simulations in
\S\ref{comparison} that use equation (\ref{Hofrho}), we set
$b_i=4/(m_iN_N$), where the constant $N_N$ is a typical {\it
initial} number of neighbors, and $h_{{\rm max},i}=+\infty$
For those one-dimensional tests that
do not implement equation (\ref{Hofrho}), we simply adjust the
smoothing lengths to always include the same number of neighbors
$N_N$, which is equivalent to enclosing the same neighbor mass as
all particles are given the same mass. For our three-dimensional
simulations, we choose $h_{{\rm max},i}= 9600R_\odot$ and $b_i$ to give
approximately the desired initial number of neighbors, either
$N_N\approx 32$ (for $N=15,780$) or $N_N\approx 48$ (for
$N=59,958$); because $h_{{\rm max},i}$ is much larger than the length scale of the problem, our results are not sensitive to its precise value.
For computational efficiency, we solve for the
smoothing lengths using a Newton-Raphson iterative scheme that
reverts to bisection whenever an evaluation point would be outside
the domain of a bracketed root.

\subsection{Pressure, Energy, and Entropy}

We associate with each particle $i$ an internal energy per unit
mass $u_i$ in the fluid at ${\bf r}_i$.  In our one-dimensional test
simulations, we implement a simple polytropic equation of state to
determine pressure,
\begin{equation}
p_i=(\gamma-1)\rho_iu_i\,,
\end{equation}
with the adiabatic index $\gamma=5/3$, corresponding to a monatomic ideal gas.
We define the total entropy in the system as
\begin{equation}
S={1\over \gamma-1}\sum_i m_i \ln(u_i\rho_i^{1-\gamma})\, .
\label{entropy}
\end{equation}

Although we need to consider only low-mass stars in
our three-dimensional calculations of this paper,
our code is quite general and includes a treatment of radiation pressure,
which can be important in more massive objects:
\begin{equation}
p_i^{\vp4}={\rho_ikT_i\over\mu_i}+{1\over 3}aT_i^4\,,
\label{pressure}
\end{equation}
where $k$ is the Boltzmann constant, $a$ is the radiation constant,
and $\mu_i$ is the mean molecular mass of particle $i$. The
temperature $T_i$ is determined by solving
\begin{equation}
u_i=\frac32{kT_i\over\mu_i}
+{aT_i^4\over\rho_i}
\,. \label{u}
\end{equation}
Equation~(\ref{u}) is just a quartic equation for $T_i$, which we
solve via the analytic solution
\citep[see, e.g.,][]{stillwell}
rather than through numerical root
finding. In particular, suppressing the subscript $i$ for convenience,
the temperature is
$$T={1\over 2}\left[-b+(b^2-4c)^{1/2}\right],$$
where 
$c=-q/(2b)+Y$, 
$b=(2Y)^{1/2}$,
$Y=Y_++Y_-$, 
$$Y_\pm=\left\{ {q^2\over 16} \pm \left[ \left({q^2\over 16}\right)^2+ \left({u\rho\over 3a}\right)^3\right]^{1/2}  \right\}^{1/3},$$
and $q=3k\rho/(2a\mu)$.
Careful programming of the expressions for $T$ and $Y_-$ can minimize roundoff error \citep[see \S5.6 of][]{numerical_recipes}.

Once the temperature $T_i$ is known, the pressure of
particle $i$ is obtained from equation~(\ref{pressure}).
For our three-dimensional calculations, we
define the total entropy in the system as
\begin{equation}
S=\sum_i m_i
\left[{3k\over 2\mu_i} \ln\left(\frac32{kT_i\over\mu_i} \rho_i^{-2/3}\right)
+ {4 aT_i^3\over3 \rho_i}\right].
\label{entropy_with_rad}
\end{equation}

\subsection{Dynamic Equations and Gravity\label{subsection:dynamic}}

To evolve the system, particle positions are updated simply by
$\dot{\bf r}_i = {\bf v}_i$ and velocities by
$\dot{\bf v}_i = {\bf \dot v}^{(\mathrm{SPH})}_i
+{\bf \dot v}^{(\mathrm{Grav})}_i$,
where ${\bf \dot v}^{(\mathrm{SPH})}_i$ and
${\bf \dot v}^{(\mathrm{Grav})}_i$
are the hydrodynamic and gravitational contributions to the acceleration,
respectively.  Various incarnations of the classical
SPH equations of motion include
\begin{equation}
{\bf \dot v}^{(\mathrm{SPH})}_i
=-\sum_j m_j
\left(
  {p_i\over\rho_i^2}+
  {p_j\over\rho_j^2}
  +\Pi_{ij}
\right) \nabla_i{\overline W}_{ij}\,
\label{classicalsph2}
\end{equation}
and
\begin{equation}
{\bf \dot v}^{(\mathrm{SPH})}_i
=-\sum_j m_j
\left(
  {p_i\over\rho_i^2}\nabla_iW_{ij}(h_i)\,+
  {p_j\over\rho_j^2}\nabla_iW_{ij}(h_j)\,
  +\Pi_{ij}\nabla_i{\overline W}_{ij}\,
\right)\, .
\label{classicalsph}
\end{equation}
The artificial viscosity (AV) term $\Pi_{ij}$ (see
\S\ref{subsection:AV}) ensures that correct jump conditions are
satisfied across (smoothed) shock fronts. The symmetric weights
${\overline W}_{ij}={\overline W}_{ji}$ are calculated as
\begin{equation}
{\overline W}_{ij}={1\over2}\Big[W_{ij}(h_i)+W_{ij}(h_j)\Big]\,.
\end{equation}

\citet{sh02} derive a new acceleration equation for the case
when an entropy-like variable is integrated.
\citet{mon02} generalized their work and showed that, even when
integrating internal energy, the same acceleration equation should
be used, namely
\begin{equation}
\dot{\bf v}_i^{({\rm SPH})} = -\sum_j m_j \left(
    {p_i\over\Omega_i^{\vp2}\rho_i^2}{\bf \nabla}_iW_{ij}(h_i)+
    {p_j\over\Omega_j^{\vp2}\rho_j^2}{\bf \nabla}_iW_{ij}(h_j)
  +\Pi_{ij}\nabla_i{\overline W}_{ij}
\right)\,,
\label{fsph}
\end{equation}
where
\begin{equation}
\Omega_i=1-{dH_i\over d\rho_i} \sum_j m_j^{\vp{\prime}}
  {\partial W_{ij}(h_i)\over \partial h_i}\,.
\end{equation}

The significant difference between equation~(\ref{classicalsph2}) or
(\ref{classicalsph}) and equation (\ref{fsph}) is the $\Omega$
factor. If the smoothing lengths do not change ($H_i=h_i={\rm
constant}$), then $\Omega_i$ is unity and equation~(\ref{fsph})
reduces down to equation~(\ref{classicalsph}), one of the many
classical variations of SPH. However, the deviation of $\Omega_i$
from unity corrects for errors that arise from a changing $h_i$, and
consequently both entropy and energy are evolved correctly.

\begin{figure}
\plotone{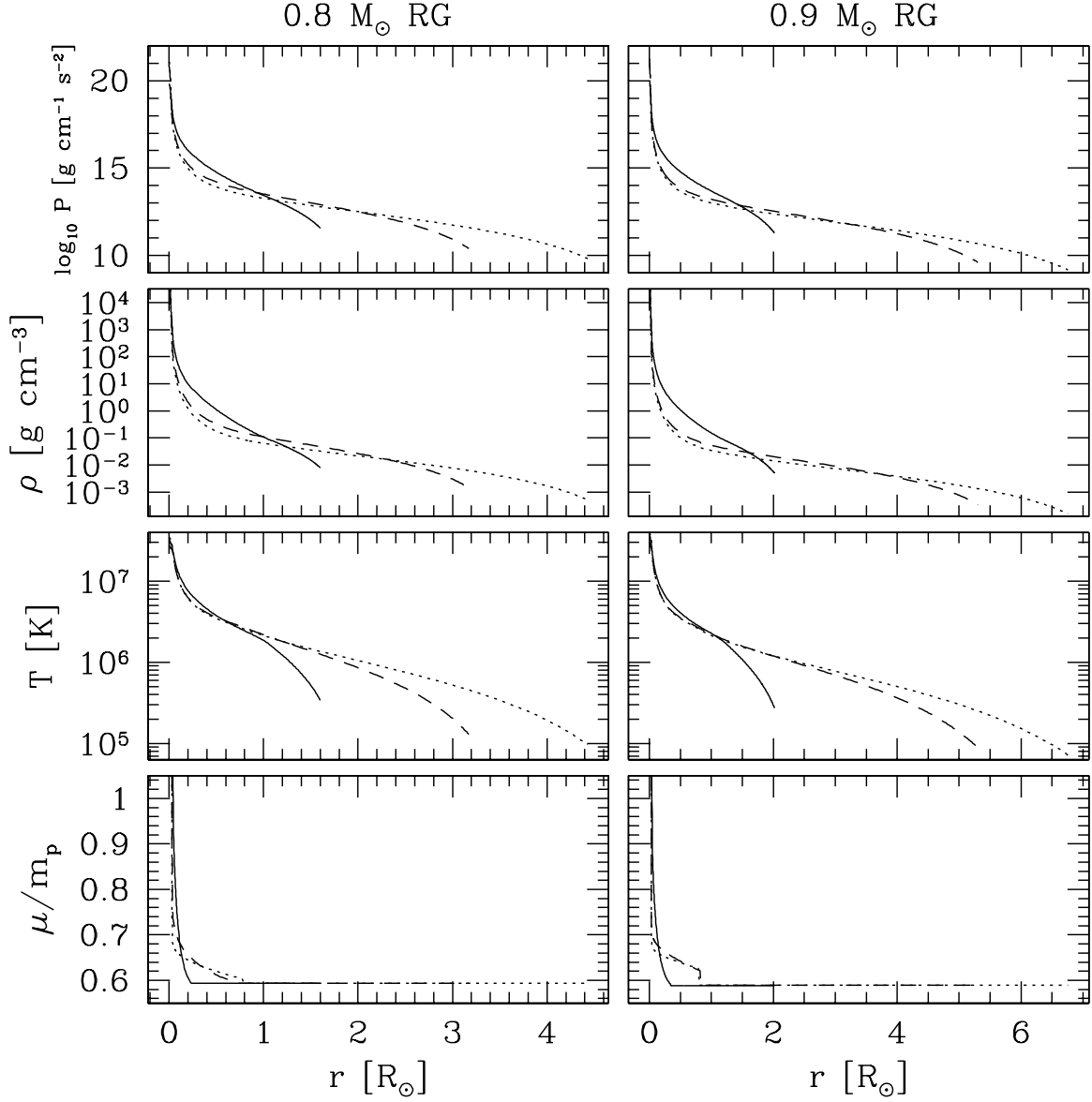} \caption{Pressure $p$, density $\rho$, temperature $T$, and
mean molecular mass $\mu$ versus radius $r$ for the parent stars with
initial masses of $0.8 M_\odot$ (left column) and $0.9 M_\odot$
(right column), as determined by our stellar evolution code. The
three curves in each plot represent the different parent stars,
from least evolved (solid line) to most evolved (dotted line).
}
\label{nataplot}
\end{figure}

We use two MD-GRAPE2 boards to calculate the gravitational
contribution to the particle acceleration as
\begin{equation}
{\bf \dot v}^{(\mathrm{Grav})}_i=-\sum_{j\ne i}
{G m_j\over r^2_{ij}+\epsilon_i^{\vp2}\epsilon_j^{\vp2}}{\bf \hat r}_{ij}\,,
\end{equation}
where $G$ is Newton's gravitational constant, $\epsilon_i$ is the
softening parameter of particle $i$, and ${\bf \hat r}_{ij}$ is the unit vector that points from particle $j$ toward particle $i$.  For the SPH particles, $\epsilon_i$ is
set approximately to the initial smoothing length $h_i$.
After trying several different values of softening parameters for the core,
we selected the one for each parent that yielded a gentle relaxation and
accurate structure profiles (see Fig.\ \ref{nataplot}).  In all cases, the
softening parameter of the core was comparable to but larger than the
largest $\epsilon_i$ in the system.  For the collision simulations, the softening parameter
of the NS is then set to that of the core.

The total gravitational potential energy is also calculated by the
MD-GRAPE2 boards, as
\begin{equation}
W=-{1 \over 2}\sum_{i,j\ne i}{G m_i m_j\over
\left(r^2_{ij}+\epsilon_i^{\vp2}\epsilon_j^{\vp2}\right)^{1/2}}\,,
\end{equation}
which clearly excludes the gravitational self energy of the NS, core, and individual SPH particles.  The GRAPE-based direct N-body treatment of gravity is crucial
for maintaining excellent energy conservation even for very long runs.
Using two MD-GRAPE2 boards, it takes about 2.5 seconds to calculate
gravitational accelerations or potentials in our $N=59,958$ simulations.

\subsection{Artificial Viscosity\label{subsection:AV}}

Tests of various AV schemes are presented by \citet{lsrs99}.
For the one-dimensional shocktube tests of this paper, we implement
the AV form proposed by \citet{1989JCoPh..82....1M} with $\alpha=\beta=1$
and $\eta^2=0.01$.
For the collision calculations of this
paper, we implement a slight variation on the form developed by
\citet{bal95}:
\begin{equation}
\Pi_{ij}=
\left({p_i\over\Omega_i^{\vp2}\rho_i^2}+
      {p_j\over\Omega_j^{\vp2}\rho_j^2}
\right)
\left(-\alpha\mu_{ij}^{\vp2}+
      \beta\mu_{ij}^2\right)\,,
\label{piDB}
\end{equation}
where we use $\alpha=\beta=1$. Here,
\begin{equation}
\mu_{ij}=
\cases{
  \medskip\displaystyle
  {({\bf v}_i-{\bf v}_j)\cdot
   ({\bf r}_i-{\bf r}_j)\over
   h_{ij}\left(|{\bf r}_i -{\bf r}_j|^2/h_{ij}^2+\eta^2\right)}\,
  {f_i+f_j\over2c_{ij}}\,,
    &if $({\bf v}_i-{\bf v}_j)\cdot({\bf r}_i-{\bf r}_j)<0$\,;\cr
  0\,,
    &if $({\bf v}_i-{\bf v}_j)\cdot({\bf r}_i-{\bf r}_j)\ge0$\,,\cr}
\label{muDB}
\end{equation}
where $f_i$ is the form function for particle $i$ defined by
\begin{equation}
f_i={|{\bf\nabla}\cdot{\bf v}|_i\over
     |{\bf\nabla}\cdot{\bf v}|_i+
     |{\bf\nabla}\times{\bf v}|_i+
     \eta'c_i/h_i}\,.
\label{fi}
\end{equation}
The factors
$\eta^2=10^{-2}$ and $\eta'=10^{-5}$ prevent numerical divergences. We
calculate the divergence of the velocity field as
\begin{equation}
({\bf\nabla}\cdot{\bf v})_i={1\over\rho_i}\sum_j
  m_j({\bf v}_j-{\bf v}_i)\cdot
  {\bf\nabla}_i W_{ij}(h_i)\,,
\label{divv}
\end{equation}
and the curl as
\begin{equation}
({\bf\nabla}\times{\bf v})_i={1\over\rho_i}\sum_j
  m_j({\bf v}_i-{\bf v}_j)\times
  {\bf\nabla}_iW_{ij}(h_i)\,.
\label{curlv}
\end{equation}
The function $f_i$ acts as a switch, approaching unity in regions of
strong compression ($|{\bf \nabla}\cdot{\bf
v}|_i>>|{\bf\nabla}\times{\bf v}|_i$) and vanishing in regions of
large vorticity ($|{\bf \nabla}\times {\bf v}|_i >>|{\bf
\nabla}\cdot {\bf v}|_i$). Consequently, this AV has the advantage
that it is suppressed in shear layers.

The only change in equation (\ref{piDB}) from Balsara's original
form is the inclusion of $\Omega_i$ and $\Omega_j$, which equal one
in simulations without adaptive smoothing lengths.  Our experience
is that the inclusion of these $\Omega$~factors within $\Pi_{ij}$
allows for a more accurate AV scheme.

\subsection{Thermodynamics}

To complete the description of the fluid, $u_i$ is evolved according
to a discretized version of the first law of thermodynamics:
\begin{equation}
{d\,u_i\over d\,t}=\sum_jm_j
\left({p_i\over\Omega_i^{\vp2}\rho_i^2}+
      {1\over 2}\Pi_{ij}\right)\,
({\bf v}_i-{\bf v}_j)\cdot{\bf\nabla}_iW_{ij}(h_i)\,.
\label{udot}
\end{equation}
We call equation~(\ref{udot}) the ``energy equation.''  A classical
form of this equation has $\Omega_i=1$ and sometimes uses the
symmetrized kernel ${\overline W}_{ij}$ in place of
$W_{ij}(h_i)$. Although \citet{mon02} uses ${\overline W}_{ij}$ for
the AV term (only), we prefer equation~(\ref{udot}) because it arises
naturally in an introduction of AV for which  $p_a/(\Omega_a\rho_a^2)$
is simply replaced by $p_a/(\Omega_a\rho_a^2)+\Pi_{ab}/2$ in both the
velocity and energy evolution equations (with $a,b=i,j$ or $j,i$).  We
find that equation~(\ref{udot}) treats shocks with essentially identical
accuracy as the Monaghan formulation, and both formulations conserve
total energy, $\sum_i m_i({\bf v}_i\cdot{\bf \dot v}_i +du_i/dt)=0$.  The
derivation of equation~(\ref{udot}) accounts for the variation of
$h_i$, so when we integrate it in the absence of shocks, the total
entropy of the system is properly conserved even though the particle
smoothing lengths vary in time.

\subsection{Integration in Time}

The evolution equations are integrated using a second-order explicit
leap-frog scheme.  For stability, the timestep must satisfy a
Courant-like condition. Specifically, we calculate the timestep as
\begin{equation}
\Delta t={\rm Min}_i
\left[\left(\Delta t_{1,i}^{-1}+
            \Delta t_{2,i}^{-1}
\right)^{-1}\right]\,.
\label{good.dt}
\end{equation}
For any SPH particle $i$, we use
\begin{equation}
\Delta t_{1,i}=C_{N,1}
{h_i\over
{\rm Max}\left[{\rm Max}_j\left(\kappa_{ij}\right),
               {\rm Max}_j\left(\kappa_{ji}\right)\right]}
\label{dt1}
\end{equation}
with
\begin{equation}
\kappa_{ij}\equiv\left[\left(
  {p_i\over\Omega_i^{\vp2}\rho_i^2}+
  \frac12\Pi_{ij}\right)\rho_i\right]^{1/2}\,,
\end{equation}
and
\begin{equation}
\Delta t_{2,i}=C_{N,2}
  \left({h_i\over
    \left|{\bf a}_i-\langle{\bf a}\rangle_i\right|}
\right)^{1/2}\,.
\label{dt2}
\end{equation}
For the simulations presented in this paper, $C_{N,1}=0.6$ to~0.9
and $C_{N,2}=0.08$ to~0.1.  The Max$_j$ function in
equation~(\ref{dt1}) refers to the maximum of the value of its
expression for all SPH particles $j$ that are neighbors with $i$.  The
denominator of equation~(\ref{dt1}) is an approximate upper limit to
the signal propagation speed near particle $i$.  The denominator
inside the square root of equation~(\ref{dt2}) is the deviation of
the acceleration of particle $i$ from the local smoothed
acceleration $\langle{\bf a}\rangle_i$, given by
\begin{equation}
\langle{\bf a}\rangle_i=\sum_j{m_j {\bf a}_j W_{ij}(h_i)\over\rho_j}\,.
\end{equation}
The advantage of including $\left<{\bf a}\right>_i$ in this way is
that the Lagrangian nature of SPH is preserved:
the timestep would be unaffected by a constant shift in acceleration
given to all particles.
For the point particles $i$ (namely, the core and the NS), we use
$\Delta t_{1,i}=C_{N,1} {\rm Min}_j\left[r_{ij}/|{\bf v}_i-{\bf v}_j|\right]$
and
$\Delta t_{2,i}=C_{N,2} {\rm Min}_j\left[r_{ij}/|{\bf a}_i-{\bf
    a}_j|\right]^{1/2}$, where the minimum is taken from all particles
$j\ne i$ in the system.  The incorporation of $\Delta t_2$ enables us
to use a larger $C_{N,1}$ value and yields an overall more efficient
use of computational resources. Indeed, we find this timestepping
approach to be more efficient than any of those studied in
\citet{lsrs99}.

\subsection{Determination of the Bound Mass and
Termination of the Calculation}

The iterative procedure used to determine the total amount of
gravitationally bound mass $M_1$ to the NS, $M_2$ to the subgiant or
red giant core, and $M_3$ to the binary is similar to that described
in \citet{lom96}. As a minimal requirement to be considered part of
a component, an SPH particle must have a negative total energy with
respect to the center of mass of that component.  More specifically,
for a particle to be part of stellar component $j=1$ (the NS) or
$j=2$ (the subgiant or red giant core), the following two conditions
must hold:
\begin{enumerate}
\item $v_{ij}^2/2+u_i-G (M_j-m_i)/d_{ij}$ must be negative,
where $v_{ij}$ is the velocity of particle $i$ with respect to the
center of mass of component $j$ and $d_{ij}$ is the distance from
particle $i$ to that center of mass.
\item $d_{ij}$ must be less than the current separation of the centers
of mass of components 1 and 2.
\end{enumerate}
If conditions (1) and (2) above hold for both $j=1$ and $2$, then the
particle is assigned to the component $j$ that makes the quantity in
condition (1) more negative.  A particle $i$ that is not assigned to
component $j=1$ or 2 is associated with component $j=3$ (the common
envelope) if only condition (1) is met for $j=1$ or for $j=2$.
Remaining particles are assigned to the component 3 if they have a
negative total energy with respect to the center of mass of the binary (mass
$M_1+M_2$), or assigned to the ejecta otherwise.

It should be noted that the method used in \citet{apjletter} to determine
mass components did not consider the possibility of a CE.
Consequently the masses reported for the binary components were larger than here,
although the conclusions of \citet{apjletter} are unaffected.

Once the mass components at a certain time have been identified, we
calculate the eccentricity $e$ and semimajor axis $a$ of the binary
from its orbital energy and angular momentum, under the approximation
that the orbit is Keplerian.  The kinetic contribution to the orbital
energy comes simply from the total mass and momentum of each component
in the center of mass frame of the binary. For the gravitational
contribution, we first use the MD-GRAPE2 boards to calculate the
gravitational energy of all fluid in the union of components 1 or 2,
subtract off the gravitational energy of just component 1, and then
subtract the gravitational energy of just component 2.  This way of 
calculating the orbital gravitational energy gives eccentricity $e$
and semimajor axis $a$ values that are much closer to being constant over an orbit than simpler methods that treated each star as a point mass.

In all cases, we wait for at
least 200 orbits before terminating a simulation. In many cases,
in which orbital properties were still varying after 200 orbits or more than a few particles were still bound to the red giant core,
we followed the evolution much longer (e.g., up to a total of 1743 orbits with
over $1.2\times 10^6$ iterations for the
case RG0.9b\_RP3.82),
in order to be certain that our orbital parameters were close to final.
For comparison, computing power limited \citet{rs91} to simulate up to only
7 orbits in their SPH study of collisions between red giants and NSs.

\section{Comparison of SPH Methods\label{comparison}}

In this section, we compare the variational SPH scheme against the
classical one for a few simple test cases.  We perform
both free expansion and shocktube tests in one dimension, problems that are
particularly useful because of their known quasianalytic solutions.
Indeed, free expansion of an initially uniform slab is just the
limiting case of the \citet{Sod1978} shocktube problem in which the
density, pressure, and internal energy density in part of the tube are
vansihingly small.  In addition, we perform a free expansion test in three dimensions.

To test the new dynamical equations (\ref{fsph}) and (\ref{udot}), we
implemented them in a one-dimensional SPH code and then compared their
results for free expansion and shocktube problems with those of runs
that use equations (\ref{fsph}) and (\ref{udot}) but with
$\Omega_i=1$ for all particles.  For each of these two formulations,
we also vary the way in which the smoothing lengths are calculated:
one run assigns $h_i$ such that each particle maintains the
desired number of neighbors, and a second one
implements the simultaneous solution of $h_i$ and $\rho_i$
by using equation
(\ref{Hofrho}).  Because we are integrating the internal energy
equation, the level of overall energy conservation is limited by the
integration technique and timestep, and not by the particular choice
of equations
being evolved: the increase in kinetic energy is always offset nicely
by the decrease in internal energy.

We begin by considering a simple case in which a uniform slab of
$\gamma=5/3$
fluid, represented by SPH particles possessing the same initial
physical properties, is allowed to expand in one dimension into the
surrounding vacuum.  This fluid has an initial density $\rho=1$ and pressure
$p=1$ over the range $-2<x<0$.
The initial specific energy values are set as $u_i=p/((\gamma-1)\rho)=3/2$,
and the system is evolved with the help of the energy equation.
In order to make a fair comparison of
various methods, we turn off the AV and implement a constant timestep.

\begin{figure}
\plotone{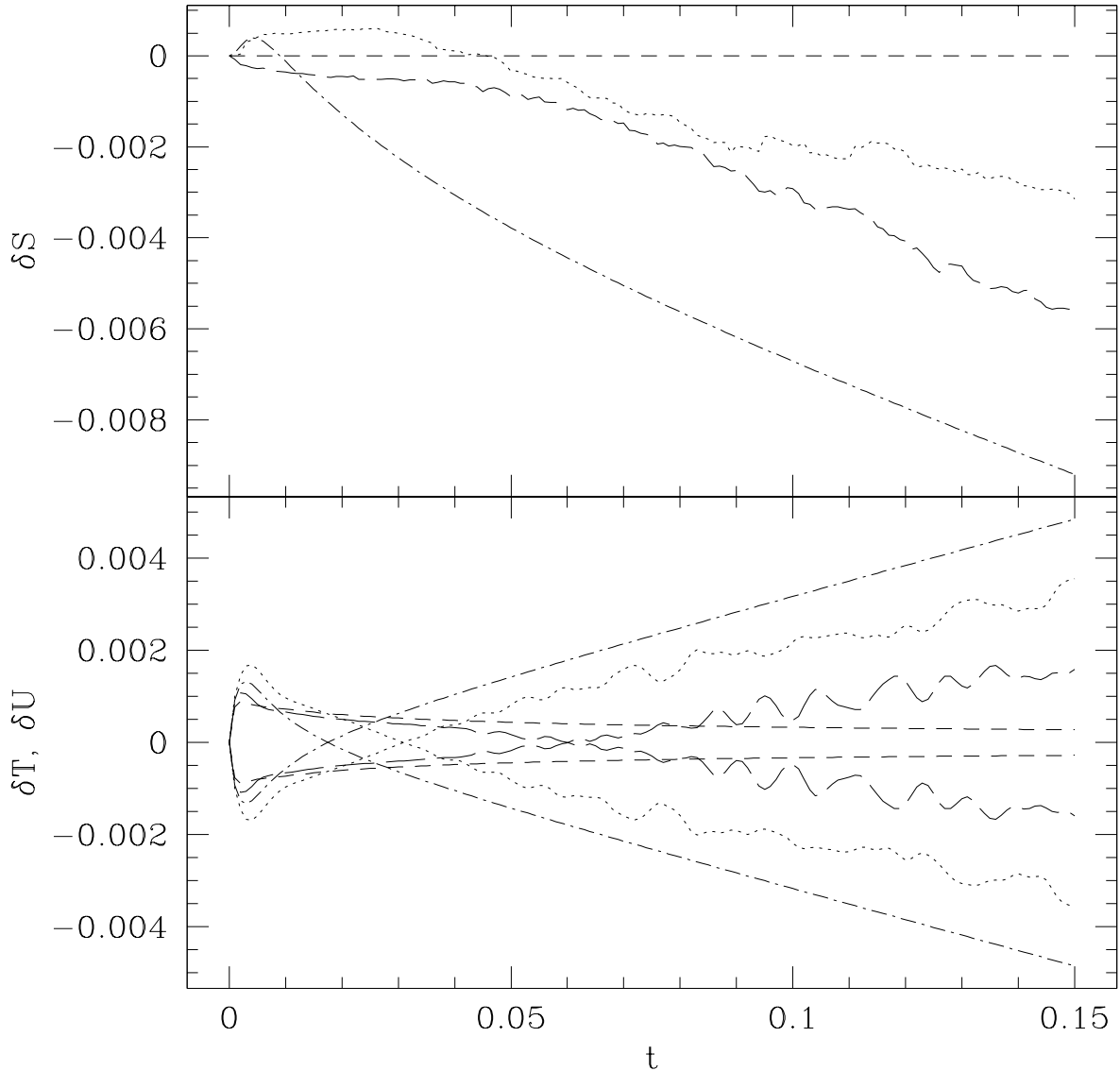}
\caption{As a function of time $t$, the errors in total entropy $S$, total
internal energy $U$, and total kinetic energy $T$ for one
dimensional free expansion calculations with $N=800$ and $N_N=5$.
Each integration method is represented by a different curve:
classical with fixed neighbor number (dotted),
classical with simultaneous solving of $h_i$ and $\rho_i$ (dot-dashed),
variational with fixed neighbor number (long dashed),
and
variational with simultaneous solving of $h_i$ and $\rho_i$ (short dashed).
In the bottom panel, the curves that are negatively valued
at early times are for $\delta T$, while the others are $\delta U$.
\label{tcfdel3_revision}}
\end{figure}

We first examine how small the errors in energy and in entropy remain
throughout the evolution.
Figure~\ref{tcfdel3_revision} presents four
calculations with $N=800$ equal mass particles, $N_N=5$, and a timestep $\Delta
t=10^{-5}$.
From the symmetry of the curves in the bottom frame of
Figure~\ref{tcfdel3_revision}, we see that all four calculations yield an error
$\delta(U+T)$ in total energy of essentially zero, which is not
surprising given that an internal energy equation is being integrated.
However 
the variational method with equation (\ref{Hofrho}) (short dashed curve) does
significantly better than any other method at keeping the errors in
total internal energy $U$ and total kinetic energy $T$ individually
close to zero.
From the top frame, we see that the classical method
(dotted curve) yields a spurious increase and then decrease in total
entropy.  When equation (\ref{Hofrho}) is used in an otherwise
classical method, the results are somewhat worse (dot-dashed curve).
Similar results
are obtained regardless of which exact form of the classical equations
are used.
Even the run
that uses the new variational equations of motion but keeps the number of neighbors
strictly fixed (long
dashed curve) produces unsatisfactory results:
entropy tends to decrease with
time. Merrily, implementing both the variational equations and
equation (\ref{Hofrho}) together (short dashed curve) produces the desired
results of essentially zero net change in energy and entropy.
Although the solution achieved by the other three approaches can be improved
by choosing larger values of $N$ and $N_N$, we find that
only the calculation
that uses the variational equations and also simultaneously solves for
$h_i$ and $\rho_i$ always has both excellent energy and entropy evolution.

\begin{figure}
\plotone{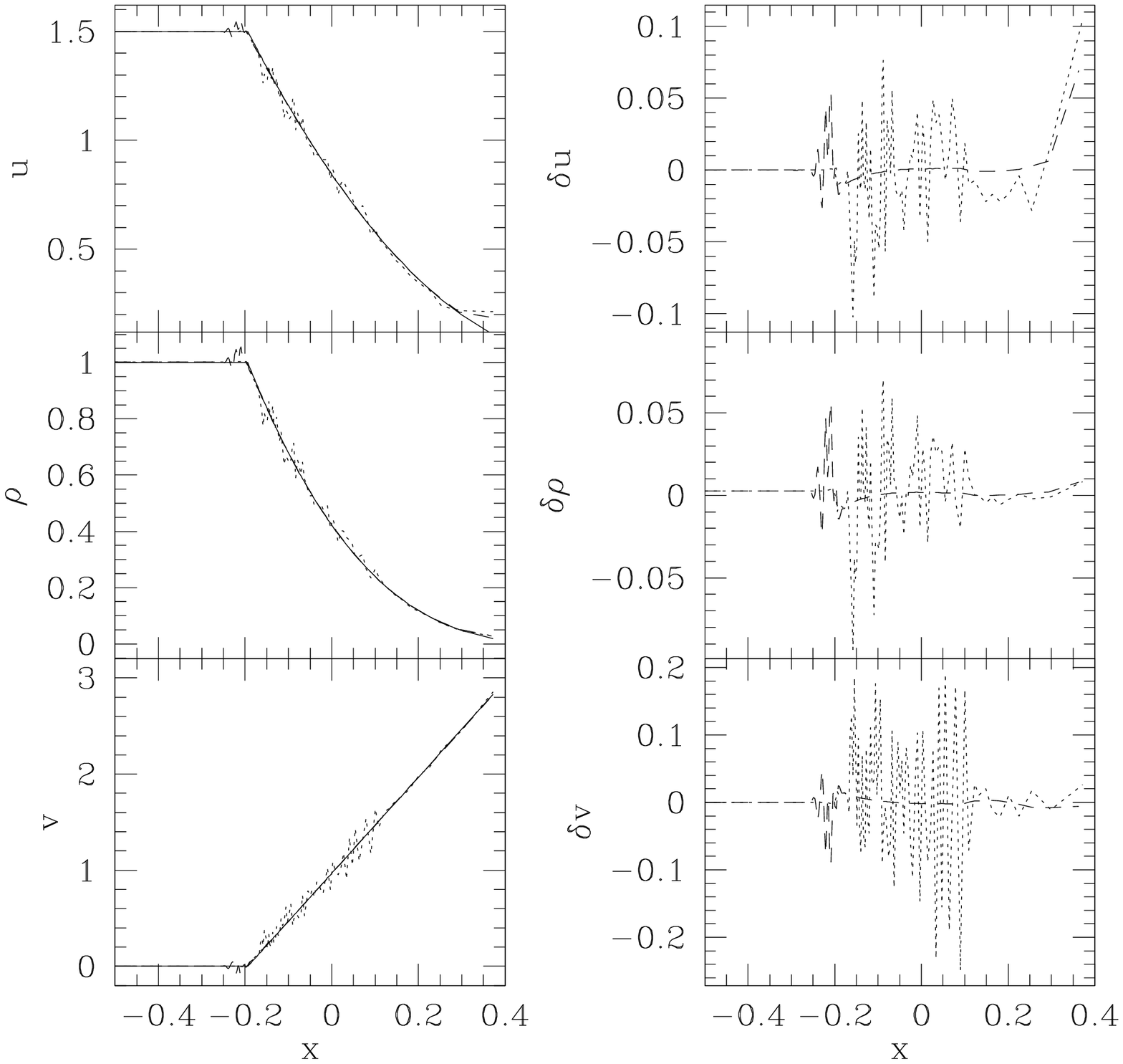}
\caption{The left column shows the specific internal energy $u$, density $\rho$, and velocity $v$
plotted as functions of position $x$ at time $t=0.15$ for two of the calculations
featured in Fig.\ \ref{tcfdel3_revision}: purely classical (dotted) and
variational with equation (\ref{Hofrho}) (dashed).
The solid curve in the left column gives the quasianalytic solution, while the right column gives the deviations from that solution.
\label{xcfdel3_final}}
\end{figure}

Figure \ref{xcfdel3_final} compares profiles of the specific
internal energy $u$, density $\rho$, and velocity $v$ against the
analytic solution for two of the calculations treated in Figure
\ref{tcfdel3_revision}: namely
 the classical method with fixed neighbor number
(dot-dashed)
and the variational method with simultaneously solution of particle
densities and smoothing lengths (dashed).
The classical calculation 
experiences significant
numerical noise, manifested in the rapid oscillation of particles.
Indeed, such oscilations are the root cause of the short timescale
fluctuations seen in Figure \ref{tcfdel3_revision} for the two
calculations that do not simultaneously solve for $h_i$ and $\rho_i$
(the dotted and long-dashed curves).
Although the variational method overshoots the analytic
solutions for $\rho$ and $u$ 
near the edge of the rarefaction wave in Figure \ref{xcfdel3_final}, it exhibits much smaller
oscillations throughout most of the fluid
and
clearly represent a much more accurate solution.  Interestingly, both
solutions overestimate the density of the unperturbed fluid at
$-1.8\lesssim x\lesssim -0.2$ by nearly 0.3\%.  This error, due to the
relatively small number $N_N$ of kernel samplings for each $\rho_i$
calculation, is the same for all of these
equally spaced particles that have not yet experienced either rarefaction
wave.

Energy conservation continues to be excellent in our tests even when
the AV is turned on.  For example when we use the same computational
parameters as above with the variational equations of motion and the
simultaneous solution of density and smoothing lengths, the total
energy changes from its initial value of 3 to 2.999999998 by the time
$t=0.15$.  As expected, energy conservation improves further when the
timestep is decreased.  The final error $\delta S$ in entropy over the
same simulation is only $9\times 10^{-5}$ (as opposed to only $3\times 10^{-8}$
when the AV is off).

Given these encouraging results with free expansion tests, we now
investigate how the variational equations and the simultaneous solution
of $h_i$ and $\rho_i$
work 
for non-adiabatic processes. In the standard \citet{Sod1978}
shocktube test, two fluids, each with different but uniform initial
densities and pressures are placed next to each other and left to
interact over time. The initial conditions for our shocktube tests are
similar to those
in \citet{rs91}, except that here the adiabatic index $\gamma=5/3$.
In particular, one slab of fluid has a denisty $\rho_l=1$ and pressure
$p_l=1$ over the range $-1<x<0$, while a second slab has
denisty $\rho_r=0.25$ and pressure
$p_r=0.12402$ over the range $0<x<1$. From the quasianalytic solution
for these initial conditions, the entropy should increase at a rate
$dS/dt=0.03834$, while the energies change at a rate $dT/dt=-dU/dt=0.9751$,
where we have included the effects both of the shock and of the free expansion at the edges.

The energy equation is integrated, with initial values of specific
internal energy set, except for the particles on the edges of the
system, according to $u_i=p\rho_i^{\zeta-1}\rho^{-\zeta}/(\gamma-1)$,
where $\rho_i$ is calculated from equation (\ref{rho}), $p$ and $\rho$
are either $p_l$ and $\rho_l$ or $p_r$ and $\rho_r$ (depending on
whether the particle is at a negative or positive $x$ position), and
$\zeta= \log(p_r/p_l)/\log(\rho_r/\rho_l)$ is chosen so that $u_i$
varies smoothly across $x=0$.  The particles near an edge (near $x=\pm
1$) are given the same $u_i$ as the bulk of the particles in that slab
of fluid.  As a result, $u_i$ is constant on each side, except over a
transition region of a few smoothing lengths near $x=0$.

\begin{figure}
\plotone{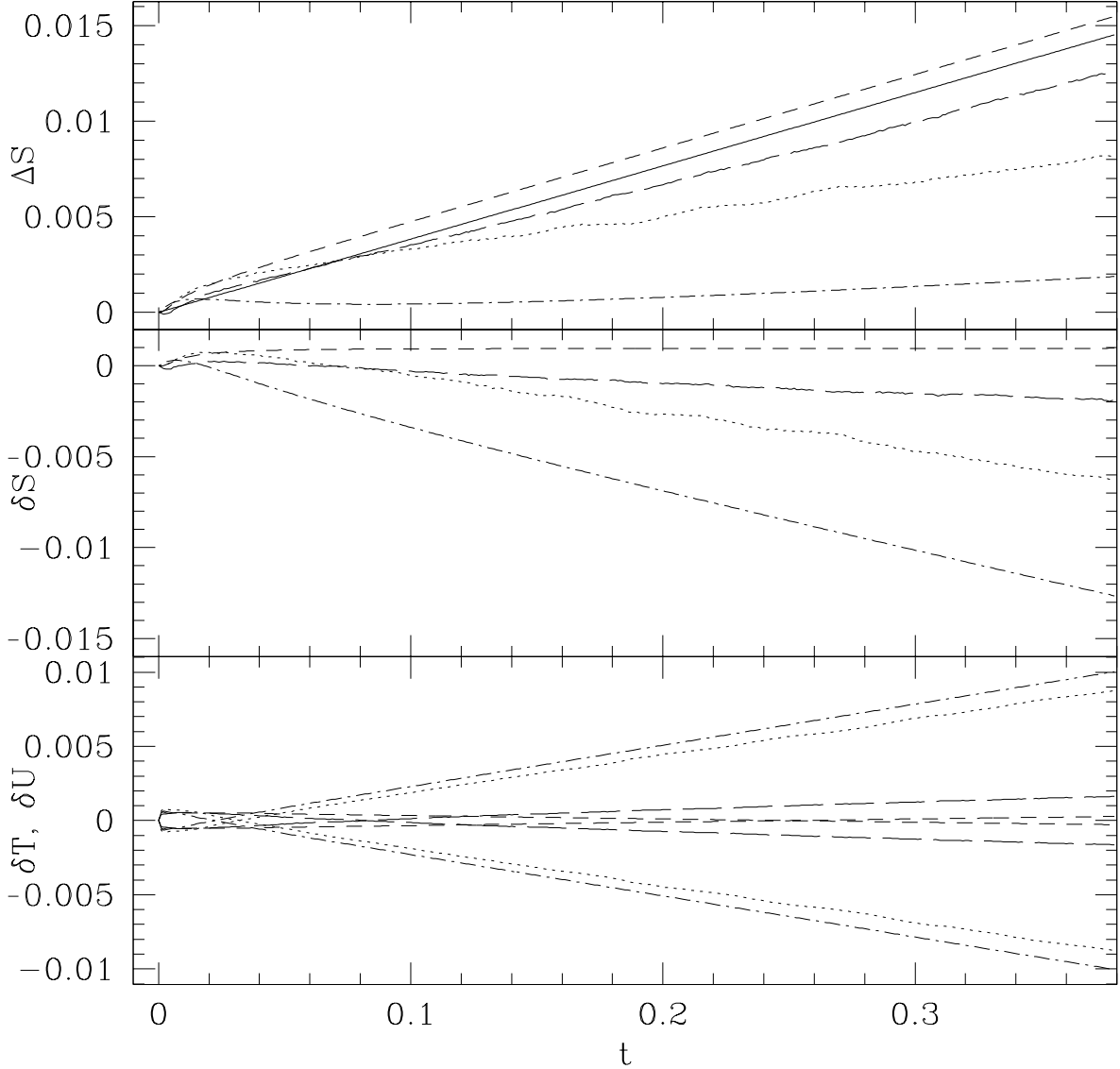}
\caption{Change in entropy $\Delta S$, error in entropy $\delta S$,
error in kinetic energy $\delta T$, and error in internal energy
$\delta U$ for a set of one dimensional shocktube test calculations
with $N=800$ and $N_N=5$.
As in Fig.\ \ref{tcfdel3_revision}, each integration method is
represented by a different curve:
classical (dotted),
classical with simultaneous solution of $h_i$ and $\rho_i$ (dot-dashed),
variational (long dashed),
and
variational with simultaneous solution of $h_i$ and $\rho_i$ (short dashed).
The solid curve in the top panel represents the
quasianalytic solution.  In the bottom panel, the
curves that are positive at early times are $\delta U$,
while the remaining curves are $\delta T$.
\label{tctdel3_revision}}
\end{figure}

Our shocktube calculations implement the same four formulations
as in the free expansion tests.
We employ $N=800$ equal mass particles
with $N_N=5$.
To help make a fair comparison of methods,
we set a constant timestep $\Delta t=10^{-5}$.
As with the free
expansion tests, we begin by examining the error in the energies and
entropy as a function of time.  From the symmetry in the curves of the
bottom panel of Figure \ref{tctdel3_revision}, we see that $\delta T=-\delta U$
to a high level of precision, as expected.  That is, all four cases
yield an error $\delta(U+T)$ in total energy of essentially zero.
We note that implementing
both the variational equations and equation (\ref{Hofrho}) together
(short dashed curve) produces the smallest errors in $T$ and $U$, a result that continues to hold even for other choices of $N$, $N_N$, and $h_{{\rm max},i}$.
From the top two panels of this figure, we see that
the entropy evolution is also most accurate when the variational equations
of motion are used and the particle densities and smoothing lengths
are found simultaneously.
The error in entropy that does remain could be further
reduced by a more sophisticated AV scheme.
The next most accurate solution uses the variational
equations with fixed neighbor number.  Neither of the two classical formulations
yield an accurate solution for these computational parameters.

\begin{figure}
\plotone{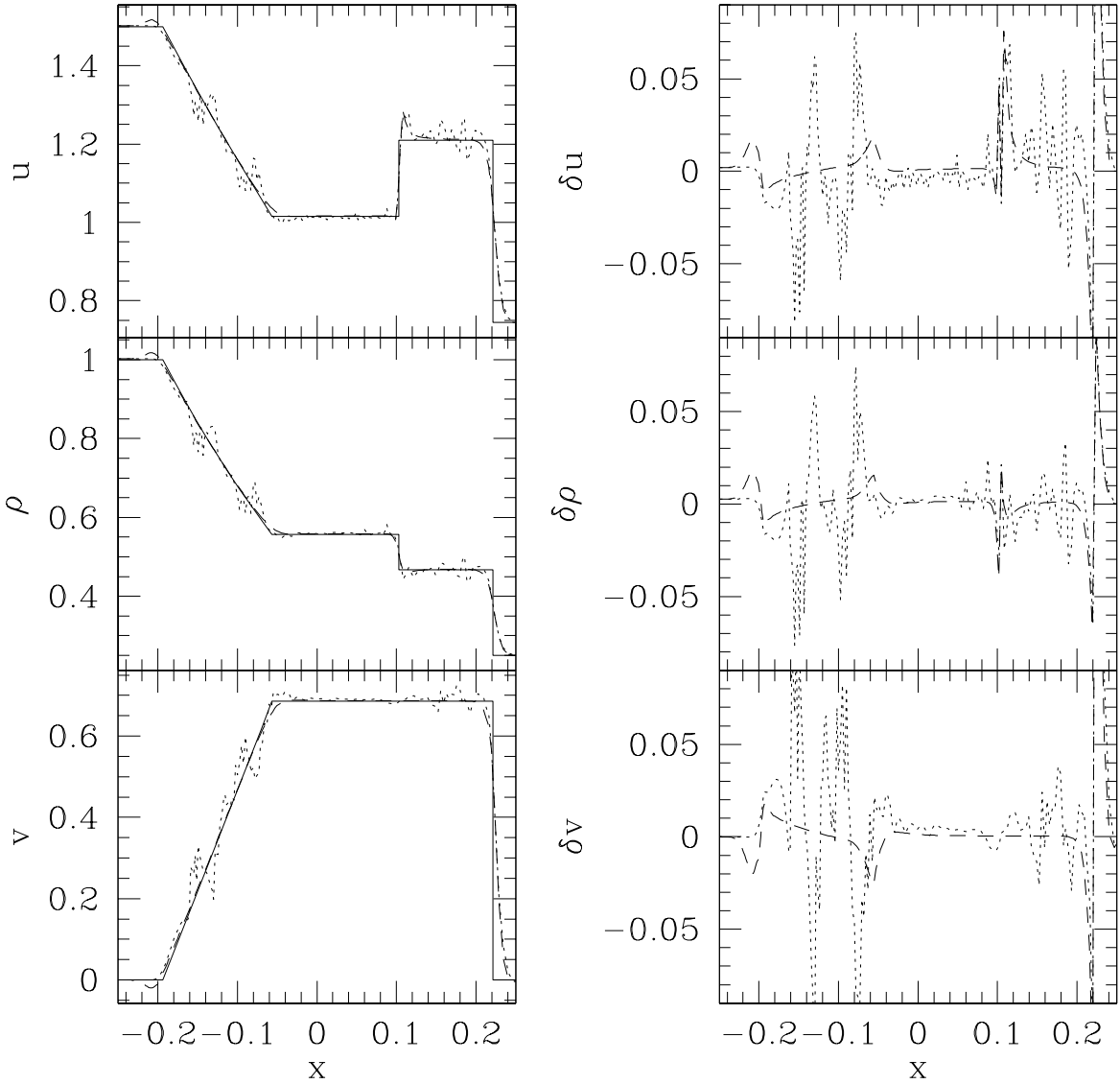}
\caption{Profiles at a time $t=0.15$ for 
two of the
one dimensional
shocktube test calculations featured in Fig.\ \ref{tctdel3_revision}: classical with fixed neighbor number (dotted) and
variational with simultaneous solution of $h_i$ and $\rho_i$ (dashed).
\label{xctdel3_final}}
\end{figure}

Figure \ref{xctdel3_final} compares profiles against the analytic solution
for the cases of the purely classical method (dotted curve) and the
variational with equation (\ref{Hofrho}) method (dashed curve).  The
purely classical method contains considerable noise, especially in the rarefaction
($-0.2\lesssim x \lesssim -0.05$). These fluctuations can be
diminished with a stronger AV, but at the expense of further
inaccuracy in the total entropy in the system.  In contrast, the
dashed curve displays very little noise or error throughout most of
the profiles, which is a direct result of solving for particle density
and smoothing length simultaneously.  Indeed, regardless of the choice of $N$,
$N_N$, and $h_{max,i}$, the variational
formulation with the simultaneous solution of $h_i$ and $\rho_i$
yields results that are as good as, or often significantly better than,
that from any other formulation considered.

\begin{figure}
\plotone{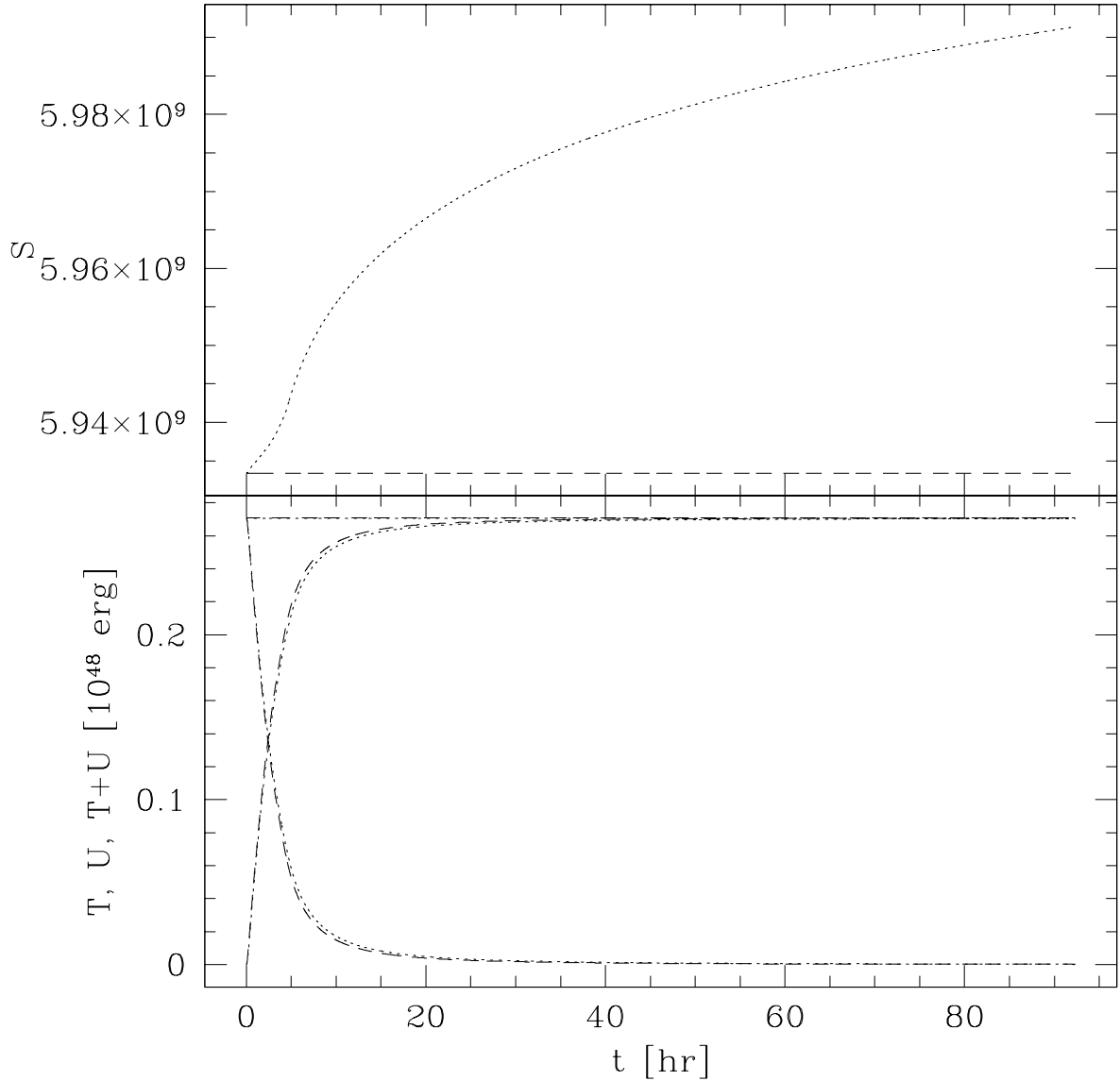}
\caption{Entropy $S$ (in cgs units), kinetic energy $T$, internal energy $U$,
and total energy $T+U$ in the
three dimensional free expansion of a $0.9 M_\odot$ red giant star after gravity is
turned off. The dotted
curve results from the classical SPH equations, while the dashed curve is
from the variational equations.  In the bottom panel, the $T$ curves increase with time, the $U$ curves decrease, and the total energy $T+U$ is essentially
constant regardless of the method.
In both calculations, equation
(\ref{Hofrho}) is used to set the smoothing lengths.
\label{fx3d}}
\end{figure}

Figure~\ref{fx3d} helps show that the SPH method in three dimensional
calculations can also be improved by the combination of the
variational equations of motion and the idea of constraining $\rho_i$
and $h_i$ analytically.  In this test, we turn off AV, set ${\bf \dot
v}^{(\mathrm{Grav})}_i=0$ for all particles $i$, use a constant
timestep of about 0.02 hours, and consider the free
expansion
of a spherically symmetric gas distribution, namely the RG0.9b red
giant model presented in \S \ref{init_data} and used in some of the
collision simulations of \S \ref{results}.
When the classical
acceleration and internal energy evolution equations are integrated,
the total entropy of the system spuriously increases with time.
Figure \ref{fx3d} demonstrates this behavior for the case where
$\Omega_i=1$ in equations (\ref{fsph}) and (\ref{udot}).
In
contrast, the variational equations allow the entropy to be properly
conserved (see the dashed line in the top panel).
\section{Modeling the Parent Stars\label{init_data}}

In this section we present our procedure for modelling the parent stars that are used in the collision simulations of \S \ref{results}.
Because the size of the NS is many orders of magnitude below our
hydrodynamic resolution, we take the usual approach of
modeling it as a point mass that interacts gravitationally, but not
hydrodynamically, with the rest of the system.
For our (sub)giant models, we first
use a stellar evolution code developed by \citet{KWH} and
updated as described in \citet{PRP2002} to evolve stars of
mass $M=0.8$ and~$0.9M_\odot$, primordial helium abundance $Y=0.25$,
and metallicity $Z=0.001$, without mass loss.
These stellar models were computed using
OPAL opacities \citep{RI} and supplemented with opacities at low
temperatures \citep{AF}. For each mass, we considered models
corresponding to different evolution stages and hence different core masses.
Table~\ref{parents}
gives the initial properties of the seven $0.8 M_\odot$ and $0.9
M_\odot$ subgiant and red giant SPH models, including one higher resolution
parent model. Column (1) gives the name of the parent star model, while
Column (2) gives its mass.  The next two columns give data resulting from the
stellar evolution calculation: column (3) gives the stellar radius $R$ of
the parent, and  column (4) lists the
core mass. The last two columns present parameters, discussed more below,
that are relevant to the SPH realization
of each model: namely, column (5) shows the mass of the central gravitational point
particle and column (6) lists the spacing of the hexagonal close-packed (hcp) lattice cells.
The models we are
considering are either subgiants (i.e., in the Hertzsprung gap)
or small red giants (i.e., near but after the base of the red giant
branch). As discussed in \citet{apjletter} and shown
in Figure~\ref{coll}, such stars are the most likely to collide.
\begin{table}[ht!]
\caption{Parent Star Characteristics}\vskip2pt
\begin{tabular}{lccccc} \tableline\tableline
Parent Star &
$M$ &
$R$ &
$m_{\rm c}$ &
$m_{\rm pt}$ &
$a_1$\\
            &
$(M_\odot)$ &
$(R_\odot)$ &
$(M_\odot)$ &
$(M_\odot)$ &
$(R_\odot)$\\
\quad(1) & (2) & (3) & (4) & (5) & (6) \\
\tableline
SG0.8a     & 0.8 & 1.60 & 0.10 & 0.15 & 0.10\\
RG0.8b     & 0.8 & 3.17 & 0.19 & 0.24 & 0.20\\
RG0.8c     & 0.8 & 4.43 & 0.22 & 0.27 & 0.28\\
RG0.8c\_hr & 0.8 & 4.43 & 0.22 & 0.25 & 0.19\\
SG0.9a     & 0.9 & 2.02 & 0.12 & 0.18 & 0.13\\
RG0.9b     & 0.9 & 5.31 & 0.23 & 0.28 & 0.33\\
RG0.9c     & 0.9 & 6.76 & 0.25 & 0.29 & 0.43\\
\tableline
\end{tabular}
\label{parents}
\end{table}

In order to generate SPH models of these subgiant and red giant stars,
we initiate a relaxation run by placing SPH particles on an hcp
lattice out to a radius of approximately two smoothing lengths less
than the stellar radius calculated by the evolution code.  Particle
masses are assigned to yield the desired density profile.
Although in general the use of
unequal mass particles can lead to spurious mixing in SPH calculations
\citep[e.g.,][]{lsrs99}, we do not expect such effects to be significant
in the collision simulations of this paper.
Because the NS is modelled as a pure point particle, there can be no
mixing of the SPH particles between the two stars.   Furthermore, the
entropy gradients in the (sub)giant parent stars and especially
in the post-collisional shocked fluid will suppress spurious mixing.

We
choose an hcp lattice for its stability \citep{lsrs99}.  Each
primitive hexagonal cell has two lattice points and is spanned by the
vectors ${\bf a}_1$,~${\bf a}_2$, and ${\bf a}_3$, where ${\bf a}_1$
and ${\bf a}_2$ have an included angle of 120$^\circ$ between them and
magnitudes $a_1=a_2$, and ${\bf a}_3$ is perpendicular to ${\bf a}_1$
and ${\bf a}_2$ with magnitude $a_3=1.633\,a_1$ \citep[see,
e.g.,][]{k86}.  In six of the parent models, we use $N=15,780$ SPH
particles and $N_N\approx 32$ to model the gaseous envelope of the
parent.  For the the most evolved $0.8 M_\odot$ parent, we also
generated a model, RG0.8c\_hr, using $N=59,958$ SPH particles and
$N_N\approx 48$ neighbors, in order to help estimate the uncertainty
in our final orbital parameters.

Because the density in the core of these parent stars is roughly $10^5$
to $10^8$
times larger than their average density, it is not
possible to model the cores with SPH particles. Instead, we model
the core as a point mass that interacts gravitationally, but not
hydrodynamically, as suggested by \citet{rs91}, among others.  The
core point mass is positioned at the origin with the six SPH
particles nearest to it located on lattice points at a distance of
$a_1/\sqrt2$.  We set the mass of the core point mass $m_{\rm pt}$ by
subtracting the sum of the SPH particle masses from the known total
mass of the star. For our parent models, $m_{\rm pt}$ ranges from
$0.15 M_\odot$ (SG0.8a) to $0.29 M_\odot$ (RG0.9c).  Because the
resolution in our parent models is much larger than the size of the core, the
core point mass $m_{pt}$ is always somewhat larger than the physical core mass $m_c$,
usually by about $0.05 M_\odot$ for our $N=15,780$ models (see Table \ref{parents}).
The increased
resolution of model
RG0.8c\_hr allows for the point mass $m_{pt}=0.25 M_\odot$
to be closer to the actual core mass $m_c=0.22 M_\odot$ given by our stellar
evolution code than it is in the RG0.8c model.

\begin{figure}
\plotone{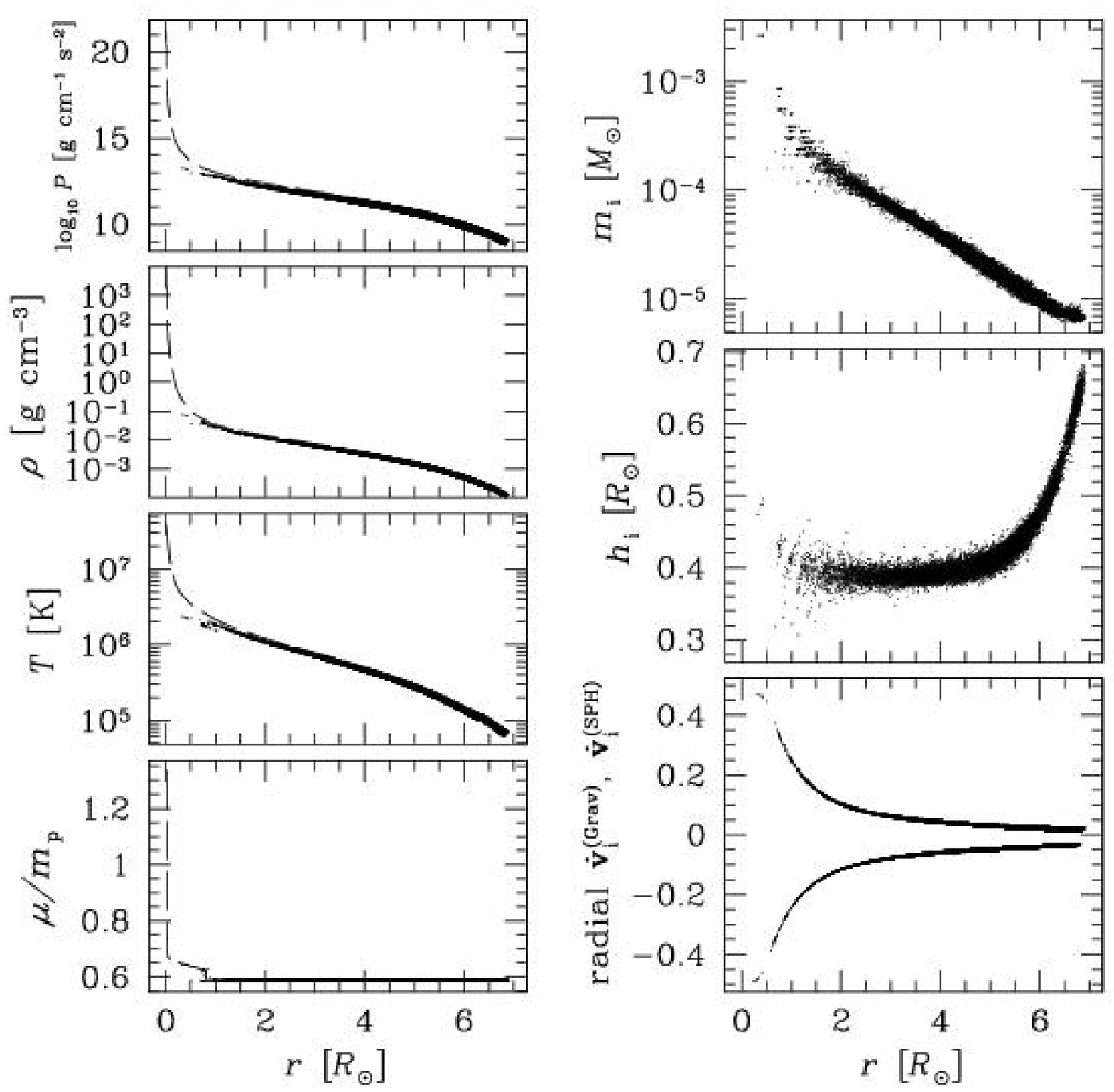} \caption{Properties of a $0.9 M_\odot$
parent star model (RG0.9c) as a function of radius.  In the left column
(from bottom): mean molecular mass, temperature, density, and
base 10 logarithm of pressure.  The dashed curves in these four
graphs represent the calculated profile of the star, while the
points represent our SPH particles.
In the right
column (from bottom): radial gravitational acceleration (lower curve) and
hydrodynamic acceleration (upper curve), smoothing length, and
individual particle mass.  } \label{RG0.9c}
\end{figure}

After the initial parameters of the particles have been assigned, we relax
the SPH fluid into hydrostatic equilibrium.  During this process, we
adjust the smoothing lengths and $b_i$ values at each iteration
in order for each particle to maintain
approximately the desired number of nearest neighbors.  We employ both
artificial viscosity and a drag force to assist with the relaxation.
We also hold the position of the core point mass fixed.
Figure \ref{RG0.9c} plots SPH particle data for one of our relaxed parent models, RG0.9c.  Although the core of the star cannot be resolved in the innermost $\sim 2$ smoothing lengths, the thermodynamic profiles of the SPH model nicely reproduce those from the stellar evolution code.

\section{Results of Collision Simulations\label{results}}

In this section, we report on the results of 32 simulations of
parabolic collisions between a subgiant or small red giant and a $1.4
M_\odot$ NS.  The results of these collisions are summarized in Table
\ref{collisions}.  In terms of computational time, most of the
$N=15,780$ particle runs lasted somewhere in the range from 5 to 20
days when run on a typical workstation with a Pentium III or IV processor.
As for the three $N=59,958$ particle runs, the $r_p=0.96 R_\odot$ case ran for
four weeks and completed its 204 orbits in nearly $7.9\times10^4$
iterations, the $r_p=1.91 R_\odot$ case lasted nearly 6 weeks and completed 219
orbits in more than $1.2\times 10^5$ iterations, and the $r_p=3.82 R_\odot$
case took 10 weeks to complete 450 orbits in about $3.25\times10^5$
iterations.  When multiple simulations were running, the time per
iteration would vary depending on the availability of the MD-GRAPE2
boards: the hydrodynamics was done on separate workstations, but
communication with the same host of the MD-GRAPE2 boards was necessary
for all calculations.

\begin{figure}
\plotone{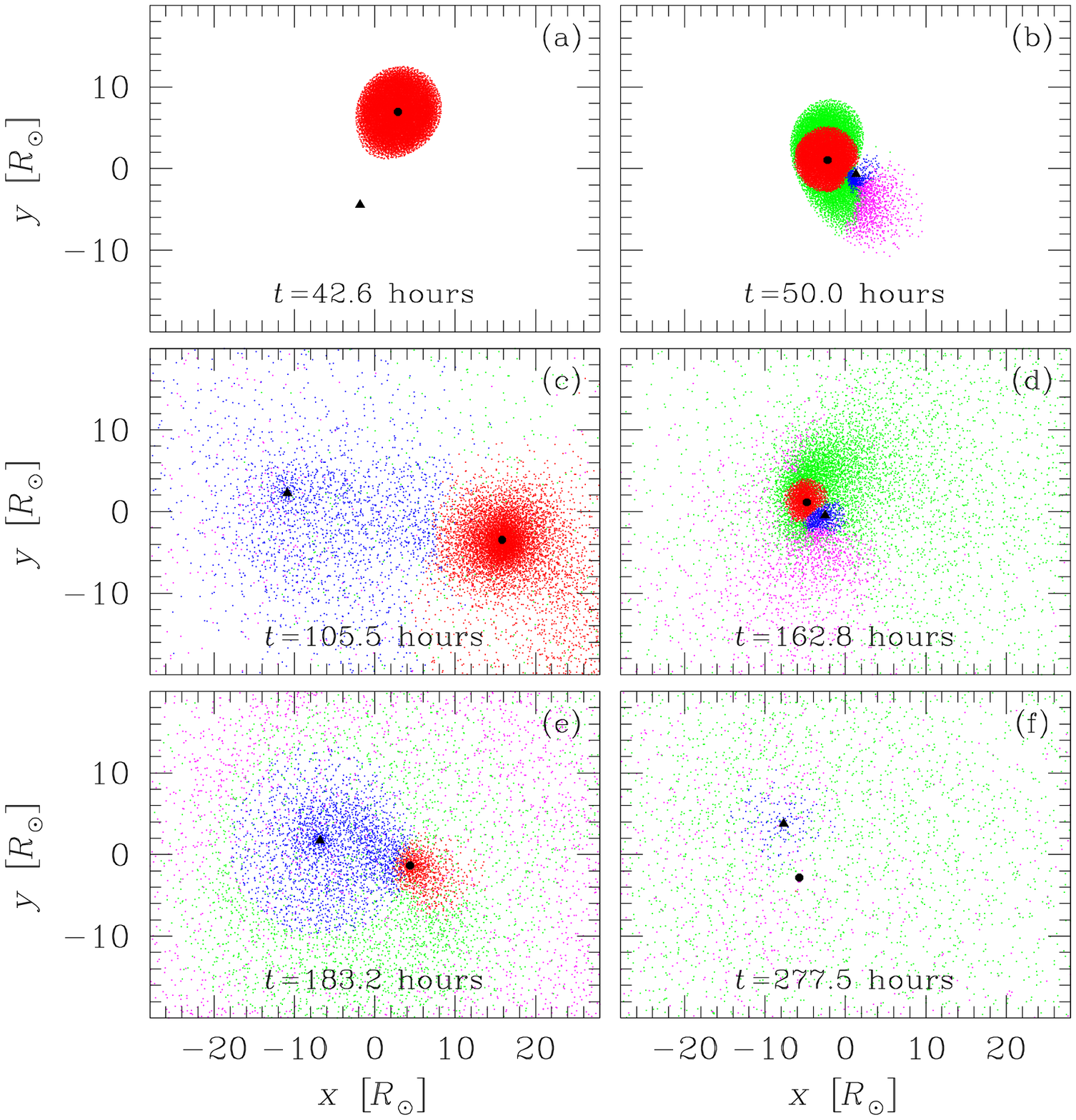} \caption{
Two-dimensional projections
onto the orbital plane of SPH particle and point-mass positions at
various times in case RG0.9b\_RP3.82.  The red giant core is
illustrated with a solid circle and the NS with a solid triangle.
Particles are colored according to the component to which they are
currently considered bound: blue for the NS, red for the red giant
core, green for the CE, and magenta for the ejecta. 
} \label{snap_9b}
\end{figure}

Figure \ref{snap_9b}
presents particle plots for one of these collisions,
case RG0.9b\_RP3.82,
in which a $0.9 M_\odot$ red giant
collides with a $1.4 M_\odot$ NS at a periastron separation of $3.82 R_\odot$ and an initial separation of $48 R_\odot$.
Frame (a) shows the two bodies just prior to impact.
Frame (b) takes place during the first
pericenter passage, while frame (c) shows the first
apocenter passage.  At this time, much of the mass
originally bound to the red giant star has been transferred to the NS.
Frame (d)
occurs during the second pericenter passage, and,
approximately twenty hours later, frame (e) occurs during the second
apocenter passage.  Finally, frame (f) shows a snapshot from the
sixth apocenter passage, at a time of 277.5 hr. At this late time,
few particles remain bound to the subgiant core.
\begin{table}[ht!]
\caption{Summary of Collisions}\vskip2pt
\begin{tabular}{lccccccr} \tableline\tableline
\quad Collision & $r_{\rm p}$ & $\Delta M_c$ & $\Delta M_{NS}$ & $M_3$ & $e$
& $a$ & $N_{\rm orbit}$ \\
 & $(R_\odot)$ & $(M_\odot)$ & $(M_\odot)$ & $(M_\odot)$ & & $(R_\odot)$ & \\
\qquad(1) & (2) & (3) & (4) & (5) & (6) & (7) & (8) \\ \tableline
SG0.8a\_RP0.24 
 & 0.24 & $0.0$ & 0.013 & 0.23 & 0.85 & 0.24 & 525 \\
SG0.8a\_RP0.48 
 & 0.48 & $0.0$ & 0.013 & 0.20 & 0.80 & 0.28 & 550 \\
SG0.8a\_RP0.96 
 & 0.96 & $0.0$ & 0.013 & 0.24 & 0.64 & 0.33 & 334 \\ \tableline

RG0.8b\_RP0.24 & 0.24 & $0.0$ & 0.005 & 0.15 & 0.89 & 0.65 & 302 \\
RG0.8b\_RP0.48 & 0.48 & $0.0$ & 0.005 & 0.11 & 0.82 & 0.77 & 235 \\
RG0.8b\_RP0.96 
 & 0.96 & $0.0$ & 0.005 & 0.11 & 0.70 & 0.98 & 207 \\
RG0.8b\_RP1.91 
 & 1.91 & $0.0$ & 0.005 & 0.10 & 0.54 & 1.26 & 259 \\
RG0.8b\_RP3.82 
 & 3.82 & $0.0$ & 0.023 & 0.07 & 0.33 & 2.28 & 1038 \\ \tableline

RG0.8c\_RP0.24 & 0.24 & $0.0$ & 0.003 & 0.15 & 0.89 & 0.95 & 201 \\
RG0.8c\_RP0.96 
 & 0.96 & $0.0$ & 0.003 & 0.09 & 0.71 & 1.34 & 281 \\
RG0.8c\_RP1.91 
 & 1.91 & $0.0$ & 0.003 & 0.10 & 0.64 & 1.78 & 298 \\
RG0.8c\_RP3.82 
 & 3.82 & $0.0$ & 0.003 & 0.06 & 0.34 & 2.19 & 862 \\
RG0.8c\_RP5.73 
 & 5.73 & $0.003$ & 0.010 & 0.05 & 0.18 & 2.70 & 1351 \\ \tableline

RG0.8c\_hr\_RP0.96  & 0.96 & $0.0$ & 0.003 & 0.08 & 0.62 & 1.15 & 204 \\
RG0.8c\_hr\_RP1.91 & 1.91 & $0.0$ & 0.005 & 0.09 & 0.51 & 1.62 & 219 \\
RG0.8c\_hr\_RP3.82 & 3.82 & $0.003$ & 0.005 & 0.08 & 0.18 & 2.18 & 450 \\ \tableline

SG0.9a\_RP0.24 
 & 0.24 & $0.0$ & 0.013 & 0.28 & 0.92 & 0.29 & 371 \\
SG0.9a\_RP0.48 
 & 0.48 & $0.0$ & 0.013 & 0.22 & 0.88 & 0.33 & 295 \\
SG0.9a\_RP0.96 
 & 0.96 & $0.0$ & 0.013 & 0.34 & 0.64 & 0.44 & 539 \\
SG0.9a\_RP1.43 
 & 1.43 & $0.0$ & 0.013 & 0.22 & 0.53 & 0.41 & 288 \\
SG0.9a\_RP1.67 
 & 1.67 & $0.0$ & 0.013 & 0.29 & 0.54 & 0.46 & 330 \\ \tableline

RG0.9b\_RP0.24 & 0.24 & $0.0$ & 0.003 & 0.18 & 0.91 & 0.97 & 297 \\
RG0.9b\_RP0.48 
 & 0.48 & $0.0$ & 0.003 & 0.11 & 0.85 & 1.07 & 531 \\
RG0.9b\_RP0.96 
 & 0.96 & $0.0$ & 0.003 & 0.14 & 0.72 & 1.26 & 365 \\
RG0.9b\_RP1.91 
 & 1.91 & $0.0$ & 0.003 & 0.12 & 0.55 & 1.65 & 434 \\
RG0.9b\_RP3.82 
 & 3.82 & $0.0$ & 0.003 & 0.06 & 0.43 & 2.12 & 1743 \\ \tableline

RG0.9c\_RP0.24 & 0.24 & $0.0$ & 0.003 & 0.16 & 0.94 & 1.41 & 261 \\
RG0.9c\_RP0.96 
 & 0.96 & $0.0$ & 0.003 & 0.14 & 0.77 & 1.67 & 213 \\
RG0.9c\_RP1.91 
 & 1.91 & $0.0$ & 0.003 & 0.10 & 0.65 & 2.08 & 200 \\
RG0.9c\_RP3.82 
 & 3.82 & $0.0$ & 0.005 & 0.11 & 0.49 & 2.79 & 325 \\
RG0.9c\_RP5.73 
 & 5.73 & $0.0$ & 0.005 & 0.10 & 0.36 & 3.17 & 416 \\
RG0.9c\_RP7.64 
 & 7.64 & $0.0$ & 0.005 & 0.12 & 0.30 & 3.37 & 303 \\ \tableline
\tableline
\end{tabular}

\tablecomments{ Final orbital properties and data for the
collisions, each labeled in col.~(1).  Col.~(2) shows the periastron
separation $r_{\rm p}$, which ranges from nearly head-on to grazing.
In col.~(3) we find the range of mass remaining bound to the
point mass representing the
subgiant or red giant core: $\Delta M_c=M_2-m_{pt}$.
Similarly,
col.~(4) displays the final mass bound to the NS:
$\Delta M_{NS}=M_1 -1.4 M_\odot$.
Col.~(5) gives the mass $M_3$ of the CE containing fluid bound to
and surrounding the final binary system. Cols.~(6) and~(7) display the final eccentricity $e$ and
semimajor axis $a$, respectively, of the resulting orbit. Col.~(8)
presents the total number of orbits followed when the calculation was
terminated.} \label{collisions}
\end{table}

\begin{figure}
\plotone{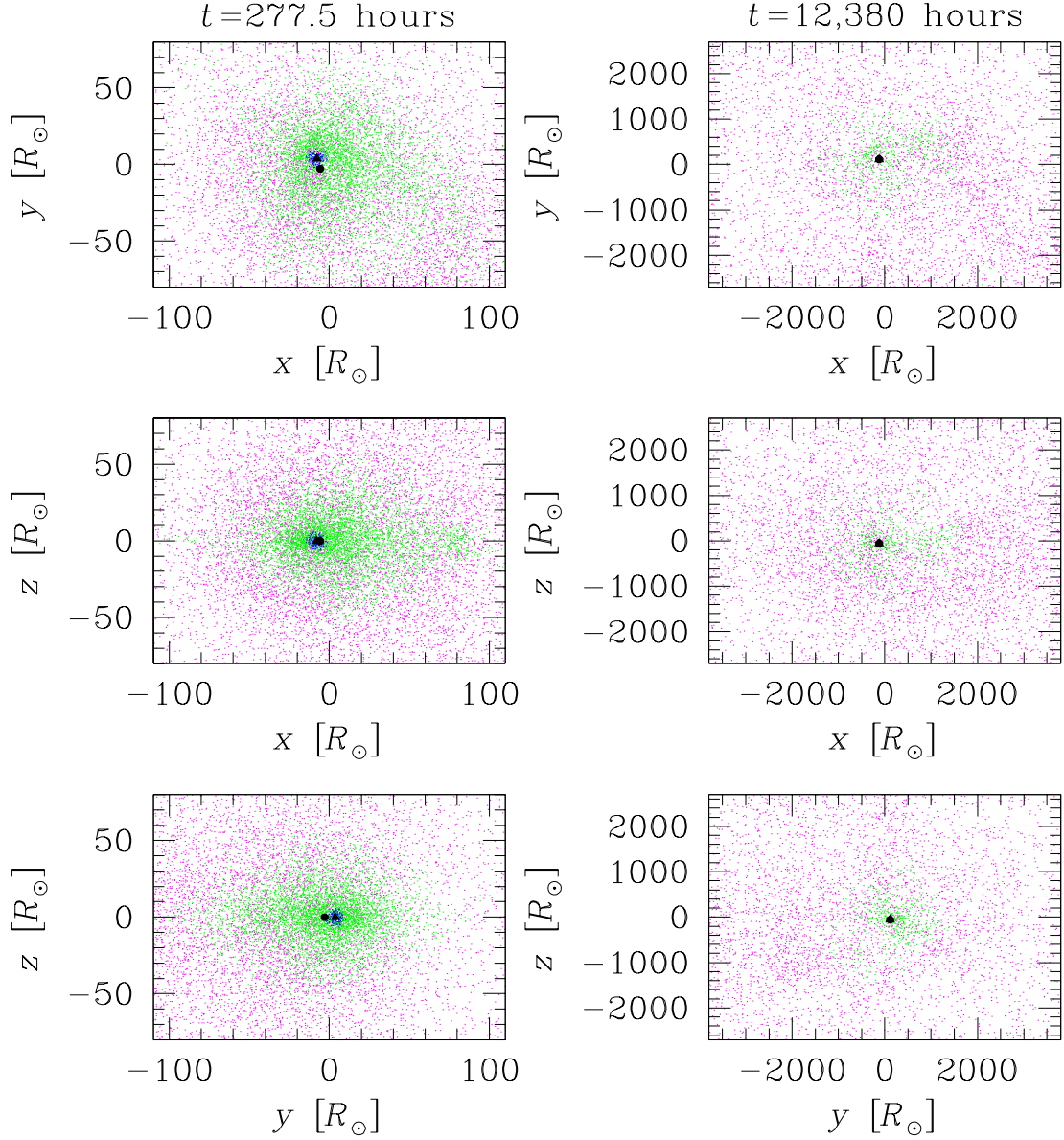} \caption{Two dimensional projections of
SPH particle and point-mass positions at two select times in the case RG0.9b\_RP3.82:
the left column is at 277.5 hr, as in frame (f) of
Fig.\ \ref{snap_9b}, and the right column at 12,380 hr.  Particle colors are as in Fig.\ \ref{snap_9b}.
} \label{snap2_9b}
\end{figure}

During all of the stellar collisions we considered,
some SPH particles form
a CE that surrounds
the binary.  Various projections of the SPH particles, including those
particles in the CE, can
be seen in Figure \ref{snap2_9b} at two different times
for case RG0.9b\_RP3.82.
At 277.5 hr, there are 158
SPH particles considered bound to the NS, 3 SPH particles bound to the subgiant core
component, and 4631 particles in the CE.  By a time of 12,380 hr, when
the simulation is terminated,
these numbers have stabilized to 1, 0, and approximately 600, respectively.
Although the outer layers of
the CE are expanding even at the end of the simulation,
this fluid is still gravitationally bound to the binary,
with a mass of approximately $0.06 M_\odot$ in this case.

\begin{figure}
\plotone{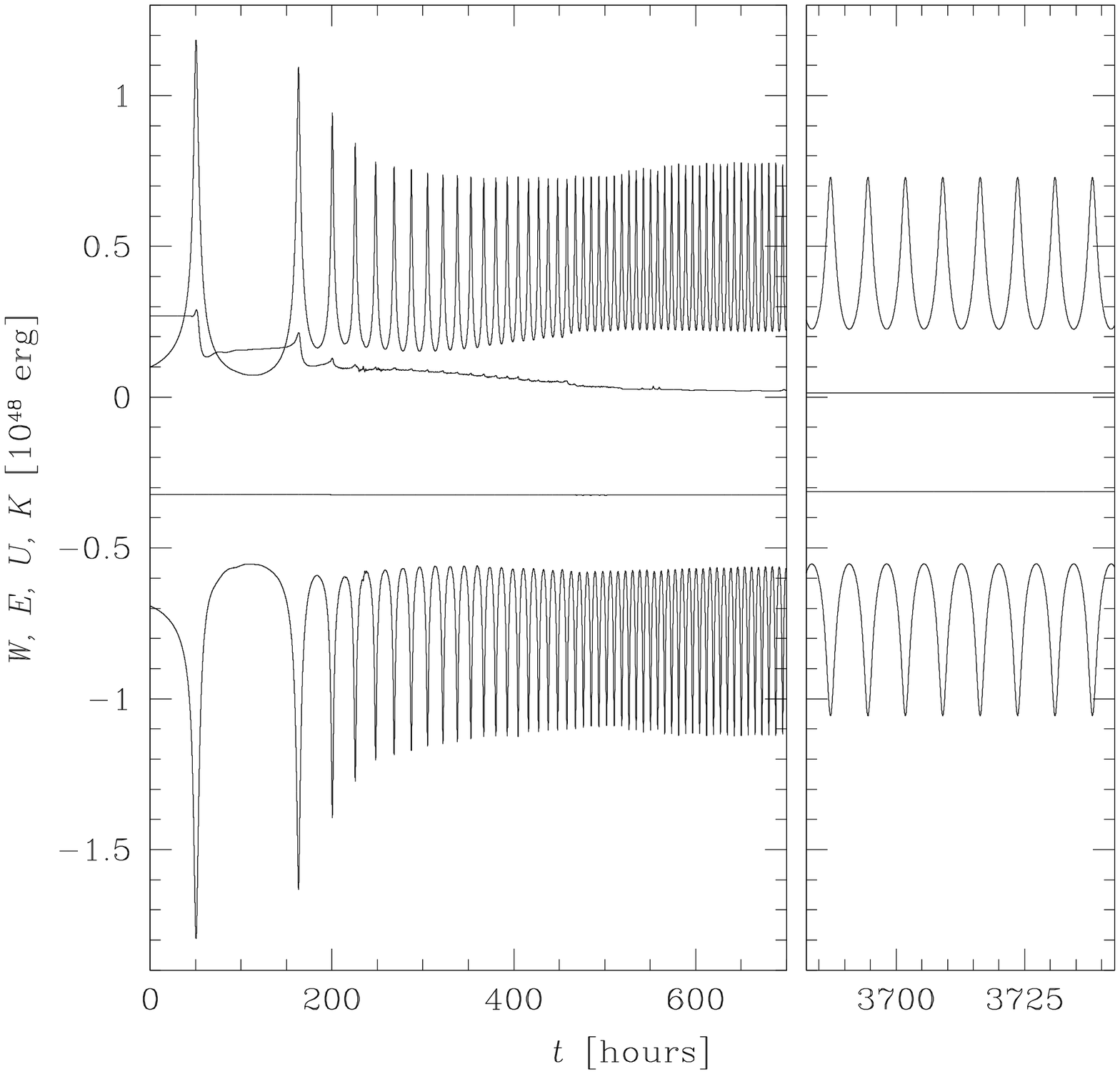} \caption {Energies as a function of time for collision
RG0.9b\_RP3.82.  The left column
presents the time evolution up to 700~hr.
The right column presents a 60 hr time interval much later in the
simulation.
The gravitational potential energy $W$ curve is the most negative, the total energy $E$ curve is nearly horizontal, the internal energy $U$ curve approaches a small positive energy at late times, and the kinetic energy $K$ curve is the positive one with large fluctuations that are synchronized with the phase of the orbit.  These energies do not include the self-energy of the core or NS.
} \label{E_9b}
\end{figure}

\begin{figure}
\plotone{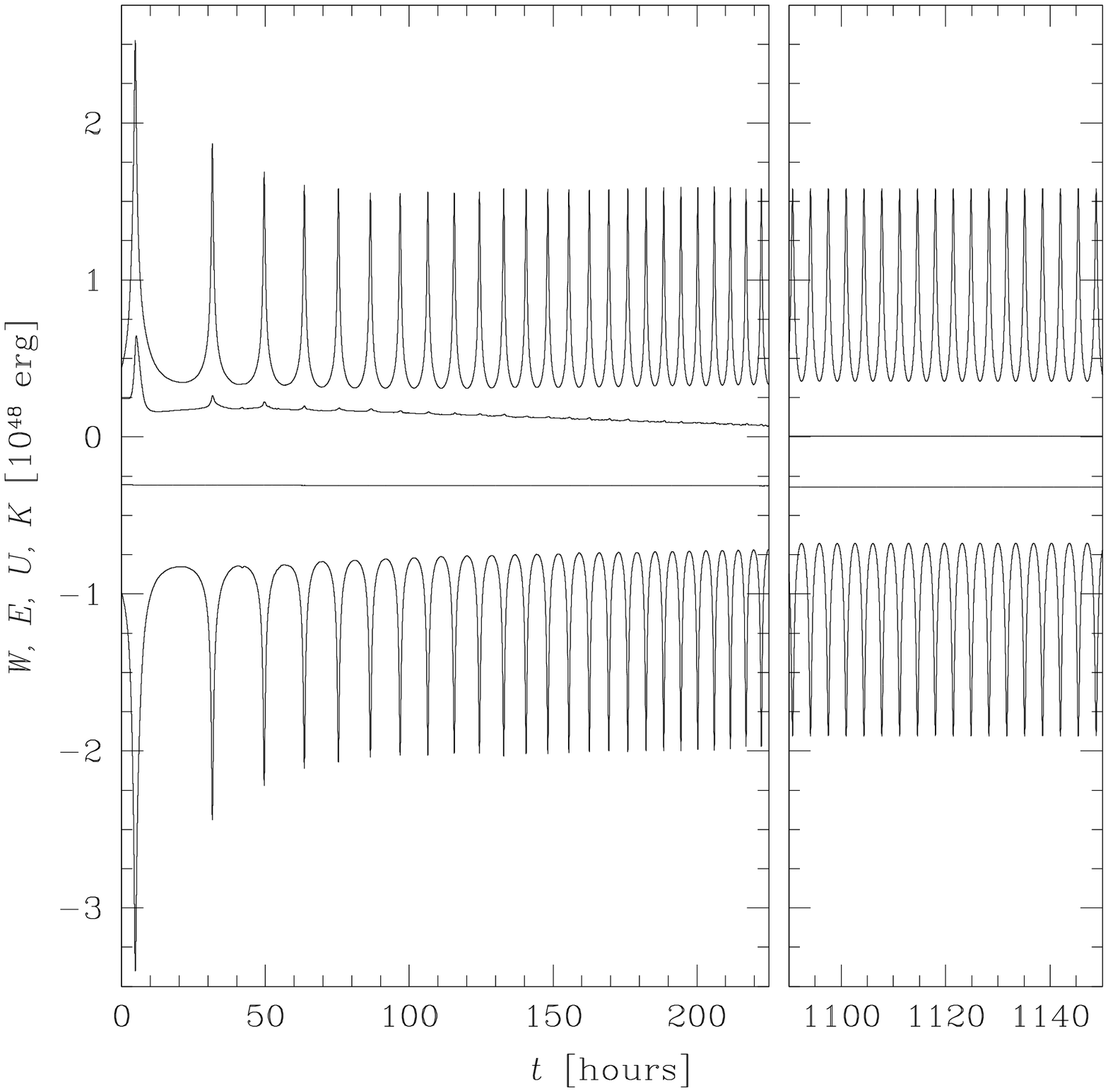}
\caption{Like Figure~\ref{E_9b}, but for the collision
  RG0.8c\_RP0.96.
}
\label{E_8c}
\end{figure}

Figures \ref{E_9b} and \ref{E_8c} show energies as a function of
time for the cases RG0.9b\_RP3.82 and RG0.8c\_RP0.96, respectively.
The sharp oscillating
peaks in the total kinetic and potential energy plots correspond to
pericenter passages, which become more frequent as the collision
progresses.
During the late times shown in the right column, the
binary has stabilized, and thus the internal energy and frequency of
kinetic and potential energy oscillations remain relatively
constant.

For our $N=15,780$ runs, the difference between the maximum and
minimum total energy is typically $\sim 10^{46}$ erg, at most a few
percent of the total energy in the system.  Energy conservation in our
$N=59,958$ runs is better by a factor of $\sim 2$.  We find that
decreasing $C_{N,1}$ and $C_{N,2}$ makes energy be conserved even more
accurately, but does not significantly affect final masses or orbital
parameters.  Energy conservation at late times, when the binary is
left surrounded by a CE, is typically excellent.  The level of energy
conservation tends to improve somewhat as we consider more evolved
parent stars or larger impact parameters.  In all of our runs, angular
momentum conservation holds at an extremely high level of accuracy,
typically at the $10^{-4}$\% level or better.  Even in the case of
RG0.8c\_RP5.73, one of our longest runs, angular momentum is conserved
to better than $3\times10^{-4}$\%.

\begin{figure}
\plotone{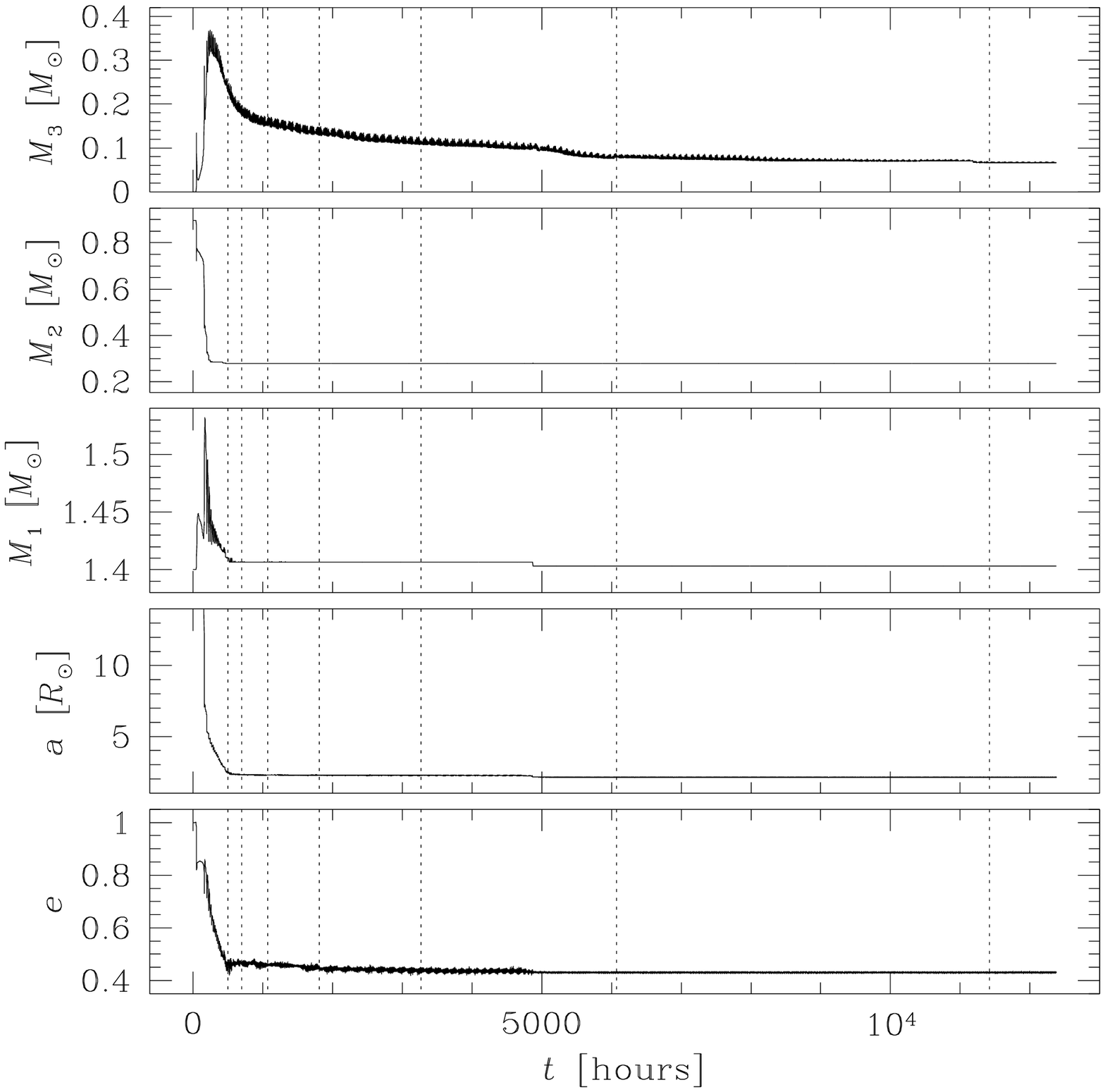} \caption{Masses $M_i$, semimajor
axis $a$, and eccentricity $e$ as a function of time for collision
RG0.9b\_RP3.82. The masses $M_1$,
$M_2$, and $M_3$ correspond to the masses containing
NS, the subgiant core, and the
CE gravitationally bound to the binary system,
respectively.  The dashed vertical lines correspond to the moments of 25,
50, 100, 200, 400, 800, and 1600 completed orbits.
} \label{newevst_9b}
\end{figure}

\begin{figure}
\plotone{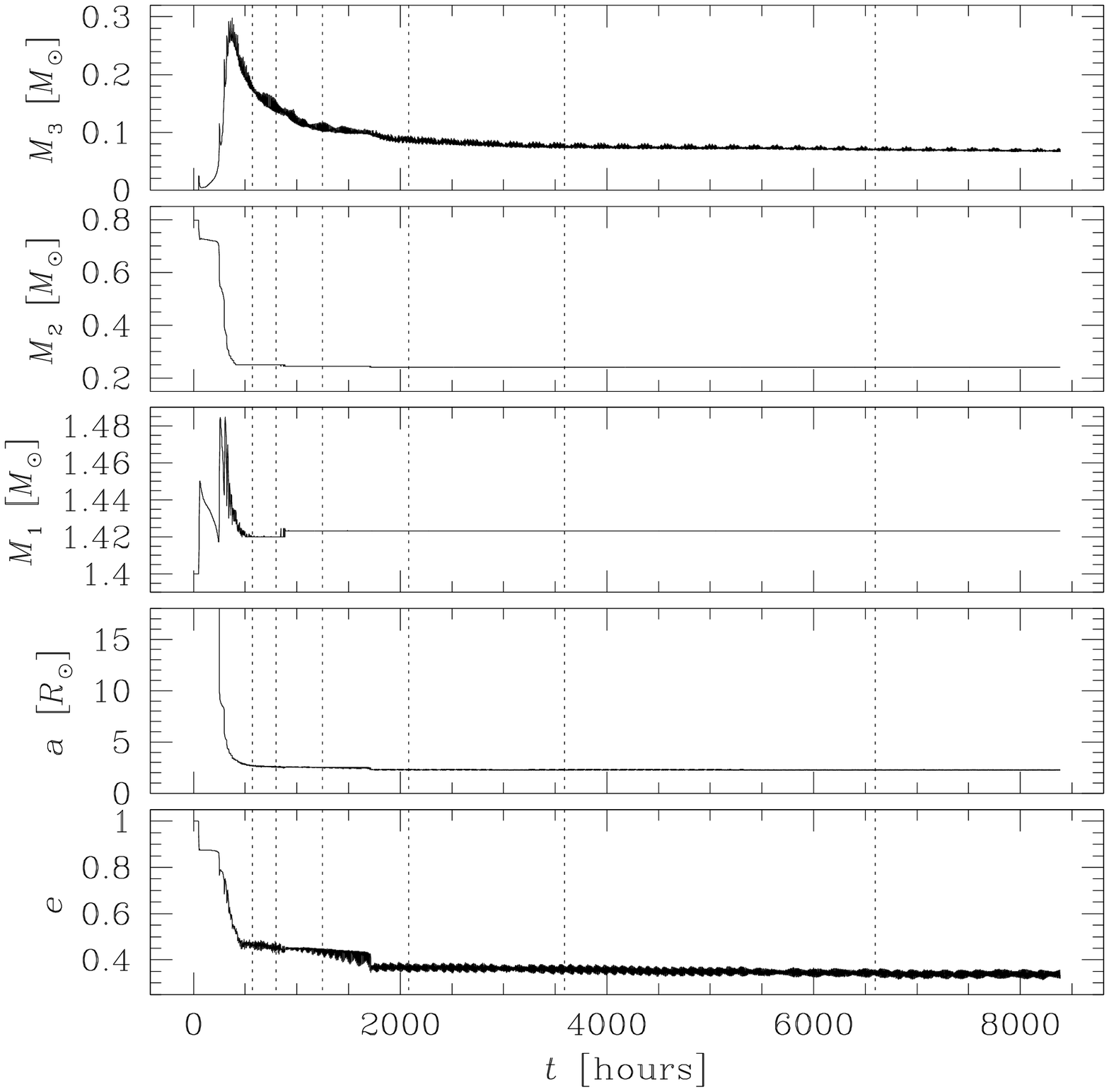}
\caption{Like Fig.\ \ref{newevst_9b}, but for the
collision
RG0.8b\_RP3.82.  The dashed vertical lines correspond to the moments of 25,
50, 100, 200, 400, and 800 completed orbits.
} \label{newevst_8b}
\end{figure}

\begin{figure}
\plotone{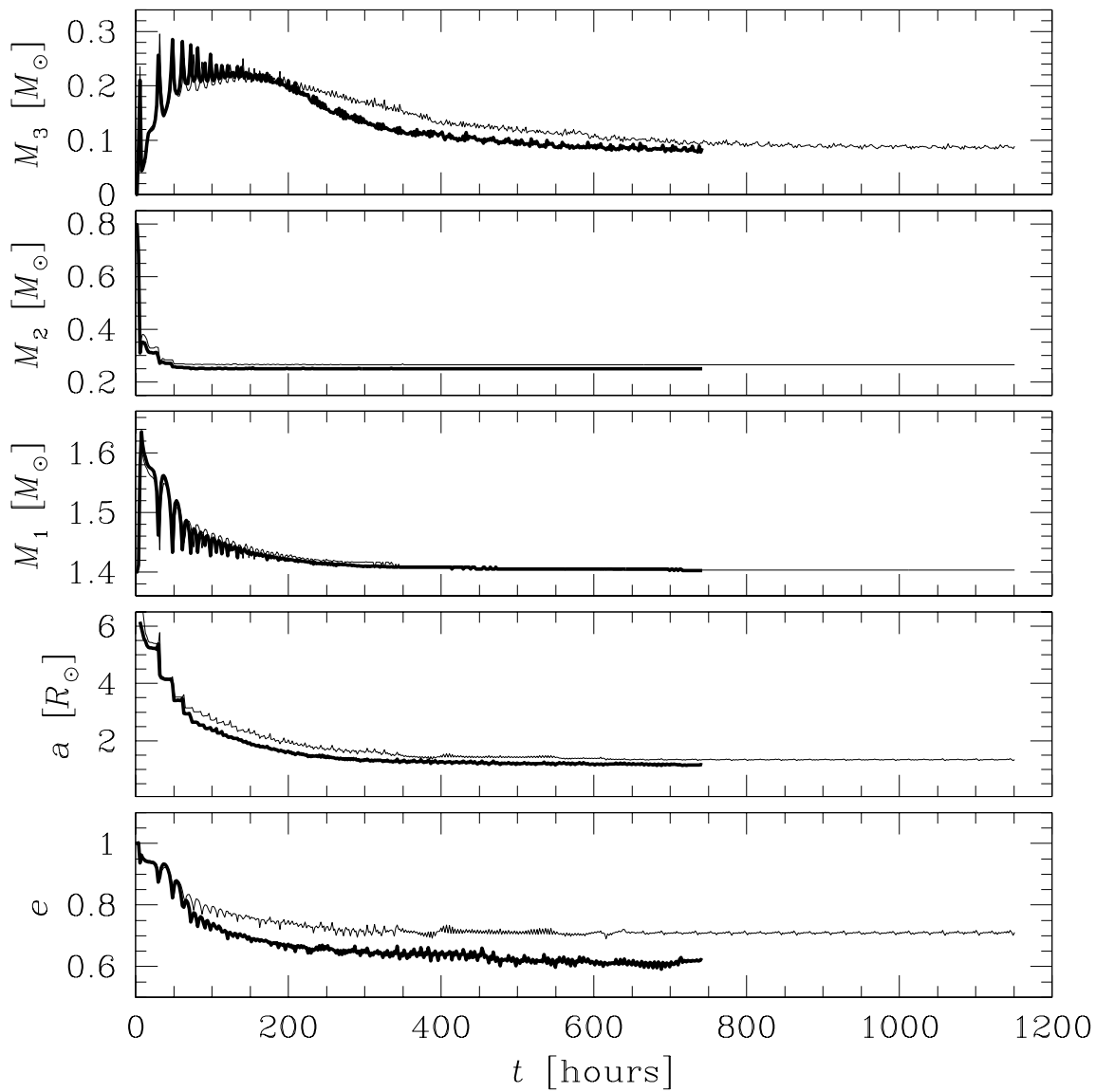}
\caption{Like Fig.\ \ref{newevst_9b}
and \ref{newevst_8b}, but for two similar collisions:
the thin curve is for RG0.8c\_RP0.96 ($N=15,780$ SPH particles and $m_{pt}=0.27 M_\odot$), while the bold curve is for RG0.8c\_hr\_RP0.96
($N=59,958$ and $m_{pt}=0.25 M_\odot$).
}
\label{newevst_8c0.96}
\end{figure}

Figure \ref{newevst_9b} shows component masses and orbital parameters
as a function of time for case RG0.9b\_RP3.82.  By 200 hours (during the third pericenter
passage), the subgiant ($M_2$) has been nearly completely stripped.  This lost mass is
transferred to the NS ($M_1$), transferred to
the CE ($M_3$), or, most commonly,
ejected to infinity.
By the end of the
collision, $M_1$ is approximately $1.403 M_\odot$ and the
eccentricity $e$ has stabilized near $0.43$.
Figure \ref{newevst_8b} shows the evolution of the component
masses and orbital parameters
for the collision RG0.8b\_RP3.82, a case that we carried out
to over 1000 orbits.  The perturbation to the orbit near a time of
$t=1700$ hours, most easily
seen in the eccentricity plot, occurs when the final gas is stripped
from the subgiant core.  After this time, the orbital parameters for
the inner binary are essentially unaffected, although there is a very
gradual decrease in the mass $M_3$ considered to be in the CE.
Figure \ref{newevst_8c0.96} shows the same parameters for the
collisions RG0.8c\_RP0.96 and RG0.8c\_hr\_RP0.96.  These two cases
both involve the same, evolved parent star, but the latter case uses
a model with more SPH particles ($N=59,958$ as opposed to $N=15,780$) and a less
massive central point particle ($m_{pt}=0.25 M_\odot$ as opposed to
$m_{pt}=0.27 M_\odot$).  In both cases, the relatively small
periastron separation leads to an eccentric orbit in which the red
giant ($M_2$) is stripped of most of the initial mass within 50 hours,
leaving only the point particle.  As the binary stabilizes, $M_1$
remains constant at $1.403 M_\odot$ in both cases.  Although the
details of the $M_3$ curves differ at early and intermediate times,
both cases settle to essentially the same final common envelope mass
$M_3$ of about $0.08$ or $0.09 M_\odot$.  In the case RG0.8c\_RP0.96,
the eccentricity $e$ tends toward $0.71$ and the semimajor axis $a$
stabilizes near $1.34 R_\odot$ at late times, while in the case
RG0.8c\_hr\_RP0.96, these quantities aproach 0.62 and $1.15 R_\odot$,
respectively.  These differences in final orbital parameters are
largely due to the different $m_{pt}$ values.  For a given total mass
and impact parameter, the trend is for smaller $m_{pt}$ to yield
somewhat less eccentric and tighter orbits: this is true regardless of
whether the $m_{pt}$ is smaller because of increased numerical
resolution (compare the higher resolution collisions of RG0.8c\_hr
models with their medium resolution counterparts) or because the
parent is in a different evolutionary stage (compare, for example,
RG0.8c\_RP0.96 and SG0.8a\_RP0.96).

\begin{figure}
\plotone{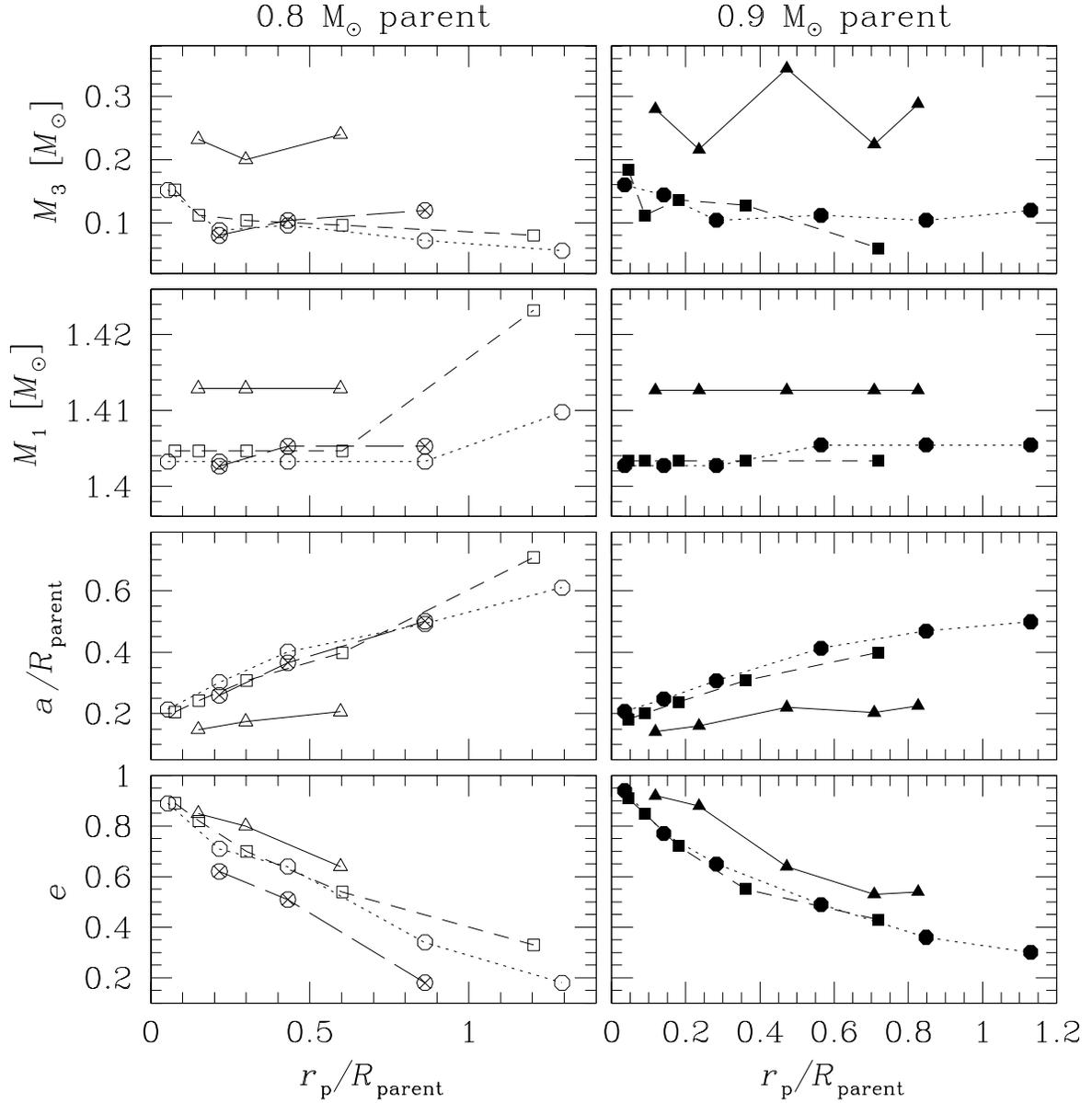} \caption{Mass $M_3$ of the CE,
mass $M_1$ of the NS and its bound fluid,
final semimajor axis $a$, and final
eccentricity $e$ as a function of normalized periastron separation
$r_{\rm p}/R$.  The left column (open shapes)
corresponds to collisions with a $0.8 M_\odot$ subgiant or giant star and the
right column (solid shapes) corresponds to collisions with a $0.9
M_\odot$ subgiant or giant star.  From least to most evolved: SG0.8a and
SG0.9a are represented with triangles, RG0.8b and RG0.9b with
squares, and RG0.8c and RG0.9c with circles.
The symbol $\otimes$ refers to collisions involving our
RG0.8c\_hr model.
} \label{masslines}
\end{figure}

By considering plots like those of Figures \ref{newevst_9b},
\ref{newevst_8b}, and \ref{newevst_8c0.96}, we are able to determine
final component masses and orbital parameters for all of our
calculations (see Table \ref{collisions} and Fig.\ \ref{masslines}).

\begin{figure}
\plotone{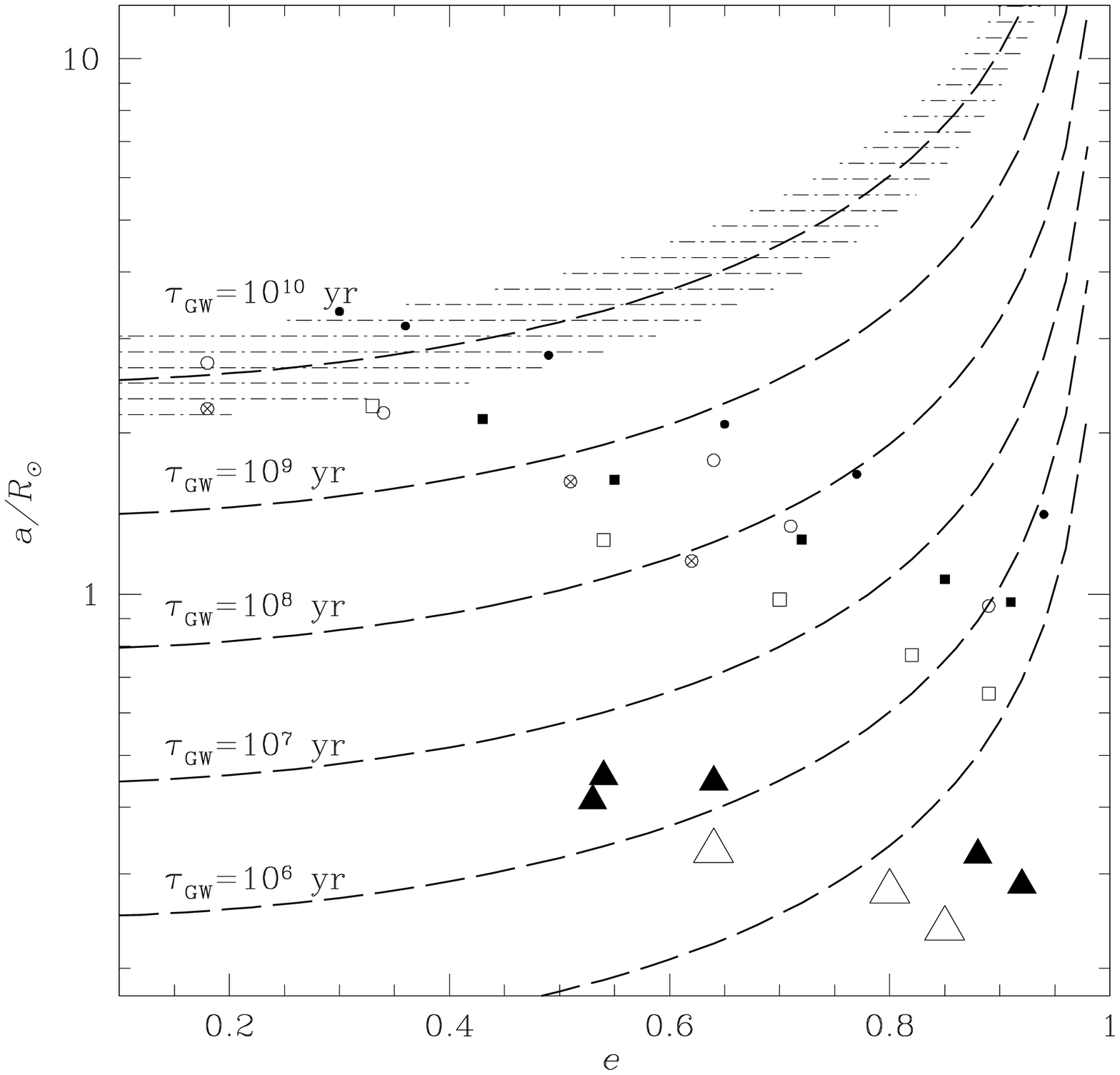} \caption{Dependence of the gravitational
radiation merger time $\tau_{gw}$ on post-collision semimajor axis $a$ and
eccentricity $e$. The points types are as in Fig.\ \ref{masslines}.
The area of the shape is indicative of $r/(dR/dt)$, where $r$ is the rate of collisions for a star of radius $R$, and hence is larger for the less evolved
subgiants, which are more likely to be involved in a collision.
The dashed curves are curves of constant gravitational merger time,
as labelled, for a 1.4 $M_\odot$ NS and
a $0.25 M_\odot$ WD.  The orbital period $P$ values on the right axis are for these same masses and are calculated simply from Kepler's third law. The hatched area shows how the
$\tau_{\rm gw} = 10^{10}\,$yr curve changes when we vary
the binary
parameters: the upper boundary corresponds to a $1.5\,M_\odot$ NS with a
$0.45\,M_\odot$ WD, and the lower boundary corresponds to a
$1.3\,M_\odot$ NS with a $0.15\,M_\odot$ WD.
} \label{avse}
\end{figure}

\section{Discussion and Future Work\label{future}}

We have shown that collisions between NSs and subgiants or small red giants
naturally produce tight eccentric NS-WD binaries that can become UCXBs.
These results reinforce our findings in \citet{apjletter}, 
where we consider the rate of UCXB formation
and show that all currently observed UCXBs could be explained from this
channel. In addition, our paper has demonstrated the importance in SPH codes
of using the variational equations of motion and simultaneously solving
for particle densities and smoothing lengths.

We have found no significant complications applying the
variational equations of motion or the simultaneous solution of $\rho_i$
and $h_i$ in either our
one- or three-dimensional hydrodynamics code.  Our simple tests of
these methods in \S3 help demonstrate the significantly
improved accuracy of SPH when
these approaches are applied together.  In particular, the
combination of these approaches produces
a change in entropy that most closely matches that of known
solutions, without disturbing energy conservation.  Additional
shocktube tests show similar results.
The reduction in
numerical noise that occurs when particle densities and smoothing lengths
are found simultaneously is due to the resulting gradual variation
of smoothing lengths from iteration to iteration.
Note that if the
smoothing lengths are instead determined by requiring that a
particle have some desired number or mass of neighbors, then these
smoothing lengths can typically take on a range of values. Noise is
then introduced by abrupt changes in the smoothing length as
neighbors enter or leave the kernel.

In our parabolic collisions of a NS and a subgiant or red giant, we
find the stripping of the core to be extremely efficient.  In all but
two cases, which were both grazing collisions with an evolved
$0.8M_\odot$ red giant, the core was stripped completely of SPH
fluid. We could not, however, resolve the innermost fluid outside the
core even in our parent models, and so the best we can do is place an
upper limit of $m_{pt}-m_c$ (usually $\sim 0.05 M_\odot$) on the
actual amount of mass that would remain bound to the core after the
collision.  As the radius of the core is about $0.02 R_\odot$ for all of
our (sub)giant parent stars, to resolve the fluid down to
the core would require an increase in the number of particles by
roughly a factor $(0.02 R_\odot/a_1)^3$, where the $a_1$ values that
set the particle spacing are given in the last column of Table
\ref{parents}.   Therefore to get the core mass correct for our least
evolved $0.8 M_\odot$ star would require about $2\times 10^6$
particles.  The other parent stars would require more particles, up to
about $2\times 10^8$ for our most evolved $0.9 M_\odot$ RG.

Some of the fluid initially in the sub giant or
red giant envelope, from 0.003 to $0.023 M_\odot$ in the cases we
considered, is left bound to the NS. This fluid will likely be
accreted onto the NS, potentially recyling it to millisecond
periods. By carrying out our collisions between a NS and a subgiant
or red giant to many orbits, we were able to identify the existence
of a residual, distended CE surrounding a binary consisting of the
NS and the subgiant or giant core. The ultimate fate of this diffuse
CE is rather uncertain. This gas will likely be quickly ejected by
the radiation released by accretion onto the NS. Nevertheless, even
a brief CE episode phase will only increase the rate of orbital
decay as compared to that from gravitational radiation alone. The
gravitational merger timescale therefore provides an upper limit on
how long it will take for a UCXB to form.

When we apply the \citet{P64} equations to these post-collision
systems, we find that most of them inspiral on rather short
timescales (Fig.\ \ref{avse}). We conclude that {\em all\/}
collisions between a subgiant and a NS, as well as all but the most
grazing RG-NS collisions, can produce UCXBs within a Hubble time. As
seen in Figure \ref{avse}, high eccentricities are an important
factor in keeping merger times short. However, as discussed in
\citet{apjletter}, even if {\em all\/} binaries were able to
circularize quickly (compared to the gravitational merger time), a
large fraction of post-collision systems would still merge in less
than the cluster age, as can be seen directly from
Figure~\ref{avse}.

If we could run all of our simulations at much higher resolution, we
would expect the details of the resulting orbital parameters to change
but our conclusions to be unaffected.  In particular, higher
resolution calculations would allow the point mass $m_{pt}$ to be
closer to the true core mass $m_c$ and could therefore more accurately
determine the mass left bound to the (sub)giant core.  Using our three
high resolution calculations as a guide, we expect that the resulting
eccentricity $e$ and semimajor axis $a$ values of the binary would also
be somewhat decreased.  However, judging from  the shape of the constant
$\tau_{gw}$ curves in Figure \ref{avse} and the change in $e$ and $a$ that
resulted by going from $N=15,780$ to $N=59,958$, we would not expect the
gravitational merger time to be dramatically affected.  That is, we
would still typically be left with a tight, eccentric binary that
would merge within a Hubble time.

\citet{BD2004} have recently shown that the cutoff in the observed
luminosity function of extragalactic LMXBs can be explained if nearly
all of the UCXB progenitors consisted of a NS accreting from a compact
object with a mass comparable to the helium core mass at turnoff.  Our
scenario for UCXB formation provides a natural explanation for such
progenitors.  When our channel for UCXB formation is followed, the
donor will be a helium WD, perhaps with small amounts of hydrogen left
over from the (sub)giant envelope.  It has long been known that the
11.4 minute UCXB 4U 1830-20 contains exactly such a helium donor \citep{1987ApJ...322..842R}.  Recently, \citet{2005astro.ph..6666I} have
also identified 2S 0918-549 as a likely UCXB with a helium WD
donor.  Although a subset of UCXBs may contain C-O WD
donors (see \citet{2005ApJ...627..926J} and references therein), 
these systems could result from
collisions between a NS and an
assymptotic giant branch star.

\acknowledgments

We thank R.\ Bi, S.\ Fleming, E.\ Gaburov, M.\ Rosenfeld, and the anonymous
referee for helpful comments and contributions.
This material is based upon work supported by the National Science
Foundation under Grants No.\ 0205991 and 0507561. N.I.~and
F.A.R.~also acknowledge support from NASA Grants NAG5-12044 and
NNG04G176G at Northwestern University. This work was also partially
supported by the National Computational Science Alliance under
Grant~AST980014N.


\end{document}